\documentclass[sigconf]{acmart}
\setcopyright{acmlicensed}
\copyrightyear{2018}
\acmYear{2018}
\acmDOI{XXXXXXX.XXXXXXX}
\acmConference[Conference acronym 'XX]{Make sure to enter the correct
  conference title from your rights confirmation email}{June 03--05,
  2018}{Woodstock, NY}
\acmISBN{978-1-4503-XXXX-X/2018/06}

\usepackage[frozencache=true,cachedir=minted-cache]{minted}

\usepackage{tcolorbox}
\usepackage[linesnumbered,ruled,vlined]{algorithm2e}
\usepackage{circledsteps}
\usepackage{soul}
\usepackage[normalem]{ulem} % Use this package for strikethrough
\usepackage{booktabs}
\usepackage{enumitem}
\usepackage{multirow} 
\usepackage{makecell}
\usepackage{subcaption}
\usepackage[page]{appendix}

\usepackage{graphicx}  % for \resizebox
\usepackage{float}     % for [H]
\usepackage{placeins} 
\usepackage{hyperref}
\usepackage[nameinlink,capitalise]{cleveref}

\newcommand{\query}{Q}
\definecolor{commentgray}{gray}{0.5}

\newcommand{\sqlin}{\textbf{\texttt{IN}}\xspace\xspace}

\newcommand{\sqlunionallcolor}{{\color{black}{{\textbf{\texttt{UNION ALL}}\xspace\xspace}}}}

\newcommand{\cntpct}[2]{#1~(#2\%)}

\newcommand{\cancut}[1]{{\color{green}{\textbf{#1}}}}

\definecolor{revone}{HTML}{1E88E5}   % Reviewer 1 (blue)
\definecolor{revtwo}{HTML}{D81B60}   % Reviewer 2 (magenta)
\definecolor{revthree}{HTML}{F57C00} % Reviewer 3 (yellow)

\newcommand{\revone}[1]{{ #1}}
\newcommand{\revtwo}[1]{{ #1}}
\newcommand{\revthree}[1]{{ #1}}
\newcommand{\revoneeq}[1]{{ #1}}

\newcommand{\ooo}[1]{{\color{orange}{\textbf{Jie: #1}}}}
\newcommand{\jie}[1]{{ #1}}

\newcommand{\cut}[1]{\sout{#1}}
\renewcommand{\cut}[1]{}

\newcommand{\inv}{\vspace*{-2mm}}
\newcommand{\sinv}{\vspace*{-2mm}}

\newcommand{\head}[1]{{{\vspace*{1mm} \noindent \textbf{#1.}}}}

\newcommand{\tool}{\textsc{GenRewrite}\xspace}

\newcommand{\lr}{\textsc{LR}\xspace}

\newcommand{\lithe}{\textsc{LITHE}\xspace}

\newcommand{\rbot}{\textsc{R-Bot}\xspace}

\newcommand{\baseline}{\textsc{Baseline LLM}\xspace}

\begin{document}

\raggedbottom
\title{\tool: Query Rewriting via Large Language Models}

% \author{Anonymous authors}

\author{Jie Liu}
\affiliation{%
  \institution{University of Michigan}
  \city{Ann Arbor}
  \state{Michigan}
  \country{USA}}
\email{jiezzliu@umich.edu}

\author{Barzan Mozafari}
\affiliation{%
  \institution{University of Michigan}
  \city{Ann Arbor}
  \state{Michigan}
  \country{USA}}
\email{mozafari@umich.edu}

\renewcommand{\shortauthors}{Liu et al.}

% ================================================

% on: with letter; off: no letter

% \twocolumn[{\centering\LARGE \textbf{Query Rewriting via LLM - Revision Letter}\par\vspace{1em}}]
% \pagenumbering{roman} % roman numerals for the letter
% \input{sections/revision_letter} 
% \clearpage                 
% \pagenumbering{arabic}

% ================================================

\begin{abstract}

Query rewriting is an effective technique for refining poorly written queries before they reach the query optimizer. However, manual rewriting is not scalable, as it is prone to errors and requires deep expertise. Traditional query rewriting algorithms fall short too: rule-based approaches fail to generalize to new query patterns, while synthesis-based methods struggle with complex queries. Fortunately, Large Language Models (LLMs) already possess broad knowledge and advanced reasoning capabilities, making them a promising solution for tackling these longstanding challenges.

In this paper, we present \tool, the first holistic system that leverages LLMs for query rewriting beyond traditional rules. 
We introduce the notion of Natural Language Rewrite Rules (NLR2s), 
which serve as hints for the LLM while also a means of knowledge transfer from rewriting one query to another, allowing \tool to become smarter and more effective over time. We present a novel counterexample-guided technique that iteratively corrects the syntactic and semantic errors in the rewritten query, significantly reducing the LLM costs and the manual effort required for verification. \revtwo{Across the standard \revthree{TPC-DS~\cite{tpcds}} and JOB~\cite{leis2015good} benchmarks and their SQLStorm-generated variants~\cite{schmidt2025sqlstorm}, \tool consistently optimizes more queries at every speedup threshold than all baselines. At the $\geq$2x threshold on TPC-DS, \tool improves 25 queries—~1.35x more than LLM-driven baselines and ~2.6x more than LLM-enhanced rule-based baselines—and the gap widens further on TPC-DS (SQLStorm); on JOB and its SQLStorm variant, where queries are simpler, absolute gains are smaller but \tool still leads by a notable margin.}

% \tool speeds up  25 out of 99 \revthree{TPC-DS~\cite{tpcds}} queries (the most complex public benchmark) by more than 2x, which is  2.5x--4.2x higher coverage than state-of-the-art rule-based query rewriting and 1.4x higher than the out-of-the-box LLM baseline.
        
        % \barzan{Jie, reaplce \% both here and in the contrib list by reporting x. For example, if we speed up 30 queries and baseline speeds up 10 queries dont say by 200\%, instead say 3x higher coverage}

% We evaluate GenRewrite against the widely recognized TPC-DS workload, demonstrating its superior performance by optimizing 22 out of 99 queries to achieve more than a 2x speedup.

% The iterative counterexample-guided correction strategy largely alleviates the burden of checking equivalence on users.

%  , GenRewrite does not discard
% incorrect rewrites. Instead, it uses 

% one sentence emphasizes the importance of query rewriting

% one sentence summarizes the limitations of exsiting rule based methods.

% in this paper, we introduce xxx empowered by the large language model

% one sentence describes the whole pipeline at high level

% evaluation results

\end{abstract}

\maketitle

\section{Introduction}\label{sec:intro}

Inefficient queries are a significant problem for nearly any organization relying on a database system. These poorly written queries—whether authored by inexperienced users or auto-generated by business intelligence (BI) tools—are a major cause of slow performance and, in cloud databases like Snowflake~\cite{dageville2016snowflake} or BigQuery~\cite{fernandes2015bigquery}, can lead to excessive costs. 
As a result, SQL-to-SQL query rewriting is a crucial step in optimizing performance and reducing expenses, 
as the likelihood of the query optimizer finding an efficient query plan is significantly diminished if it is given a poorly-written query to begin with. 
However, whether performed manually or automatically, query rewriting has remained a challenging task.

\head{Challenges of manual query rewriting}
The poorly-written queries that have the biggest impact on the overall cost and performance tend to be quite complex too. 
For example, it is not uncommon for queries auto generated by
Looker~\cite{lookergoogle} to span 100s of lines.
Ensuring the rewritten query preserves the semantics of the original requires a deep understanding of the database schema, constraints, and the user intent, and is therefore a time-consuming and error-prone task, even for seasoned database experts.
The fact that there are often thousands, if not millions, of such slow queries makes the manual approach a monumental undertaking.

\head{Limitations of traditional automated query rewriting} 
Despite extensive research, automated query rewriting techniques have long faced fundamental limitations that hinder their practical effectiveness. The most common approach relies on \emph{rewrite rules}—--pattern-matching transformations that replace query fragments with optimized equivalents while preserving semantics. 
Regardless of whether the rules are developed and provided by experts~\cite{bruno2022computation,pirahesh1992extensible,ahmed2006cost} or 
are inferred automatically~\cite{wang2022wetune,ding2023proving},
a rule-based query rewriter is as effective as the set of rules provided to it. 
By definition, rule-based query rewriters\footnote{Rule-based query \emph{rewriting} should not be confused with rule-based query \emph{optimization}; the latter is simpler, and thus much more common in practice~\cite{graefe1987rule}.}
% are limited to situations where the input query matches one of their existing rules,  
% and thus
are fundamentally incapable of optimizing new query patterns that do not match their existing rules~\cite{dong2023slabcity}.
Although synthesis-based query-rewriting does not require prior rules~\cite{dong2023slabcity}, in practice it only handles simple queries---for instance, SlabCity~\cite{dong2023slabcity} optimizes just 2 out of 22 TPC-H~\cite{tpch} queries and 3 out of 99 TPC-DS~\cite{tpcds} queries.

\head{Large language models (LLMs)} The shining success of large language models (LLMs) in performing complex and open-ended tasks has created new hopes for solving some of the hardest database problems as well. 
While researchers have used LLMs for Text-to-SQL~\cite{fu2023catsql,gao2023text,sun2306sql,katsogiannis2023survey,pourreza2024din} and a few other areas, e.g., data cleaning and integration~\cite{narayan2022can,suhara2022annotating}, table processing~\cite{li2023table,lu2024large} and database diagnosis~\cite{zhou2023d}, their potential for query rewriting has remained largely unexplored. In instances where human experts or existing rules cannot rewrite a query, say for complex queries or new query patterns, 
 could LLMs be leveraged to uncover rewrite opportunities due to their broad general knowledge and advanced reasoning capabilities? If so, this could drastically lower the burden on human experts and allow for a much larger set of queries to benefit from rewriting.
To the best of our knowledge, this paper is the first\footnote{See \S\ref{sec:related} for related work published after the initial publication of our work on ArXiv.} to focus on 
 answering this question: how to best leverage LLMs for query rewriting beyond rules and how to address the challenges involved.

\head{Challenges} Using LLMs for query rewriting has challenges:\footnote{A common misunderstanding is that the time and monetary overhead of invoking an LLM may outweigh the speedup benefits of the rewrite. However, this one-time overhead is less relevant for repetitive workloads, which are the focus of \tool (see \textbf{Target workloads}).} 
\begin{enumerate}[label=C\arabic*, leftmargin=0.5cm]

\item The na\"ive  approach of simply prompting the LLM to "rewrite the given query into a faster but equivalent form"  can rarely handle complex queries: 
most outputs either fail to parse (syntactic errors) or produce incorrect results (semantic errors) due to SQL’s complexity and LLM hallucinations (\S~\ref{sec:expr:llmbaseline}).
%~\cite{zhang2023siren,tonmoy2024comprehensive}.

\item LLMs lack a database cost model and cannot run experiments, so their rewrites are not necessarily faster than the original queries.

\item While LLM costs for rewriting a recurrent workload can be amortized, excessive invocations can still become expensive.
     
\item Hints can help the LLM, but excessive or irrelevant ones can lead to mistakes.

\item Due to their black-box nature, LLM rewrites are harder for humans to understand and verify than traditional  rules.

\end{enumerate}

%ta inja
\head{Our approach}
To address the above challenges, we present \textbf{\tool}, the first holistic system that leverages LLMs for query rewriting. 
We introduce the notion of a Natural Language Rewrite Rule (NLR2), which is a textual explanation summarizing the rewrite. 
We use the LLM itself to produce the NLR2s, which are then used for three purposes. The NLR2s 1)  serve as hints assisting the LLM to provide better rewrites 
(C2), requiring fewer follow-up LLM invocations (C4), 2) make the rewrites easier to understand and verify by offering a human-readable explanation (C5), and, 
most importantly, 3) allow \tool to transfer the knowledge gained from rewriting one query to another and thus become smarter and more effective over time (C1, C2, C3). Since NLR2s are void of confidential information (such as column or schema names), one can even bootstrap \tool's NLR2 repository for a new workload using the repository of a previously trained \tool on another workload.

To avoid redundant rules, which can in turn confuse the LLM, we maintain
    a utility score for each NLR2 to supply only those rules to the LLM  as hints that are most relevant to the query at hand.
Finally, to minimize the LLM cost and minimize the human effort needed for manual verification, 
\tool does not discard incorrect rewrites. Instead, it uses an novel counterexample-guided technique to iteratively
    correct the syntactic and semantic errors in the rewritten query (C1, C3).

\begin{figure}[ht]
\inv\sinv
  \centering
  \includegraphics[width=\linewidth]{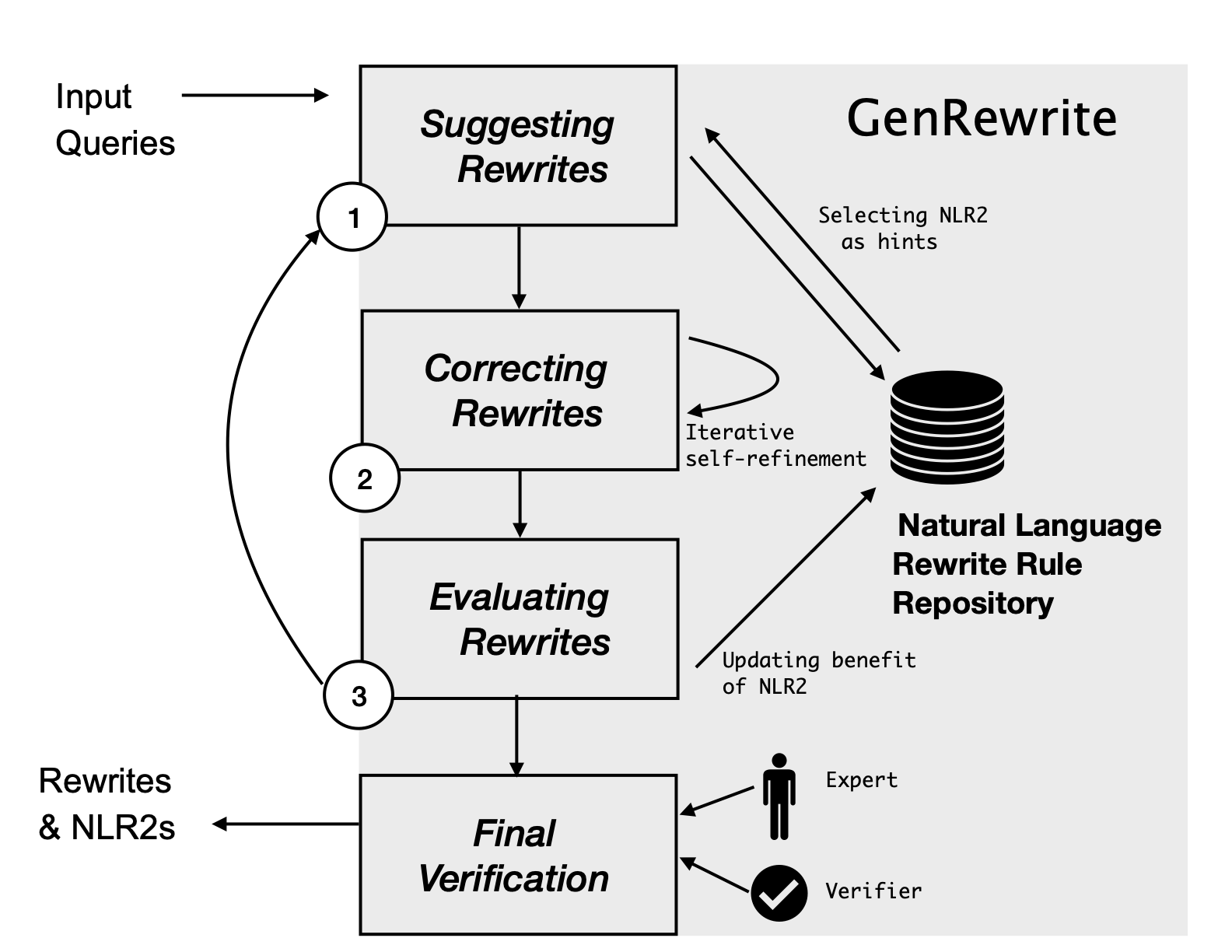}
\inv\inv\inv
  \caption{High-level Workflow of \tool }
  \label{fig:arch}
\inv
\end{figure}

% \barzan{don't forget to fix the plot. sent you a bunch of feedback on this yesterday} \jie{btw it is really easy to get lost in voice msgs on whatsapp, could you please remind me of the problems of this figure if you still remember}

\head{High-level workflow} 
\tool takes as input a workload,\cancut{\footnote{A streaming setting is a special case, whereby queries are presented and rewritten one at a time.}}
expressed as a set of input queries $\mathcal{Q}$, and returns
\{($q$,$q'$,$e$) \textbar  $q$ $\in$ $\mathcal{Q}_{opt}$ \}, where
  $\mathcal{Q}_{opt}\subseteq \mathcal{Q}$ is the optimizable subset of the input queries, 
  $q'$ is the rewritten query that is equivalent to $q$ but likely to run faster, and $e$ is 
 a human-readable explanation of the rewrite.
Figure~\ref{fig:arch} shows the overall workflow, which is an iterative process.
In each iteration, every query in the workload goes through three phases to find a better rewrite than the one found in previous iterations:
{\small \Circled{1}} rewrite suggestion, {\small \Circled{2}}  rewrite correction, and {\small \Circled{3}}  rewrite evaluation. 
Additionally, \tool \revthree{maintains} a Natural Language Rewrite Rule (NLR2) repository for knowledge transfer across queries in the workload.
{\small \Circled{1}} selects appropriate NLR2s from the repository to serve as hints for suggesting new rewrites. Any newly discovered NLR2 is also added to the repository. 
{\small \Circled{2}} iteratively refines rewrites based on the counterexamples provided by the LLM  to eliminate semantic and syntactic errors.
  {\small \Circled{3}} evaluates the rewrites for equivalence and performance, and updates the benefit score of the used NLR2s accordingly.
This process continues until no additional improvement is found and the set of rewrites stabilizes. 
The rewrites then go through a final verification step {\small \Circled{4}}, 
where if the formal verifier cannot guarantee equivalence automatically,
a human expert is asked to manually inspect and confirm  the rewritten query's equivalence before it is used or deployed to production.

% \jie{yes, this module asks for counterexamples; I will mention it. The output from (2) is what LLMs believe are correct. There may be a gap between what LLMs believe are correct and what are actually correct. So we can add a correctness checking module in (3). (3) checks equivalence and speedups at the same time.}

\head{Target workloads} Repeatedly invoking an LLM incurs a time and monetary overhead. 
% For instance, \tool uses a default budget of 45 seconds that can be adjusted by users. 
 As mentioned above, when the formal verifier cannot guarantee the equivalence, a human expert needs to inspect the rewrite, which incurs an additional delay.
 For these reasons, \tool is only focused on the following use cases:
\begin{enumerate}[leftmargin=0.5cm]
\item \textbf{Recurrent workloads}, where the same query template is executed many times, such as BI (Business Intelligence) dashboards, reporting, ETL, DB-based applications, and dbt~\cite{dbt} jobs. 
For instance, 
over 60\% of the jobs in Microsoft clusters are recurrent~\cite{jindal2018selecting}.
Here, queries can be rewritten and optimized by \tool once and then reused many times.

\item \textbf{Highly expensive queries} expected to take 10s of minutes, e.g., an adhoc query scanning terabytes of data.
Here, even if the query is run once, the additional overhead of leveraging an LLM or a human expert is still well justified.
\end{enumerate}

 In other words, \tool is \emph{not} designed for real-time adhoc workloads, where queries are novel \textbf{\underline{and}} relatively light, as the additional overhead of an LLM cannot be amortized.

\head{Contributions} We make the following contributions:
\begin{enumerate}[leftmargin=0.5cm]
    \item We present one of the first in-depth analyses of LLMs for query rewriting, outlining key challenges and opportunities for future research in this area (\S\ref{sec:insight}).
    \item We present \tool, a holistic tool leveraging LLMs for autonomous query rewriting. It 
        addresses the key challenges of using LLMs by introducing (i) the notion of Natural Language Rewrite Rules (NLR2) for finding better rewrites and effective knowledge transfer,
          (ii)  a utility score for NLR2s to minimize LLM confusion caused by irrelevant or too many rules , and (iii) a counterexample-guided iterative correction method.
        (\S\ref{sec:approach}).
    \item We evaluate \tool on the standard TPC-DS and JOB benchmarks and \revtwo{their SQLStorm-generated variants}. \tool consistently optimizes more queries at every speedup threshold than all baselines. \revtwo{At the $\geq$2x threshold on TPC-DS, \tool improves 25 queries—~1.35x more than LLM-driven baselines and ~2.6x more than LLM-enhanced rule-based baselines—and the gap widens further on TPC-DS (SQLStorm); on JOB and its SQLStorm variant, where queries are simpler, absolute gains are smaller but \tool still leads.} (\S\ref{sec:expr}).
       
\end{enumerate}

\section{Motivation and Background}
\label{sec:motivation}

\begin{comment}
    
\subsection{Recurring Resource-Intensive Queries} 

\jie{ Enterprise data analytics often consists of recurring queries over evolving data, which, despite being essential to daily operations, can be highly resource-intensive. In fact, over 60\% of the jobs in Microsoft clusters have been reported to be recurrent~\cite{jindal2018selecting}. Since such queries often run on a fixed schedule (e.g., nightly or multiple times per day), suboptimal SQL logic can accumulate into significant performance bottlenecks and increased cloud spend over time. Thus, identifying, analyzing, and optimizing these heavy queries is critical to sustaining an efficient data workflow.}

\jie{Real-world dbt case studies~\cite{dbtcasestudy1,dbtcasestudy2} illustrate how data teams manually refactored complex SQL queries—executed four times per day—by removing redundancies, streamlining joins, and leveraging incremental models—efforts requiring deep domain knowledge and focused engineering expertise. Although these refactorings often dramatically reduce query runtimes, the process is nontrivial and labor-intensive. Consequently, an automated and interpretable solution is needed to help systematically address these recurring performance bottlenecks.} 

\ooo{Add a problem definition here for our query rewriting setting}

\end{comment}

\subsection{Limitations of Traditional Query Rewriting}

Despite decades of effort by experts to craft extensive rewrite rules, many queries still cannot be optimized by existing rules~\cite{dong2023slabcity}. 
In those cases, leveraging general knowledge becomes essential in identifying redundant calculations in the query and exploring alternative ways of achieving the same output. 
To demonstrate this, we present two examples: one from LeetCode~\cite{leetcode}, the other from TPC-DS~\cite{tpcds}, an industry-standard business intelligence benchmark.

\begin{figure}[h]
  \centering
  \begin{minipage}{.45\textwidth}
    \begin{minted}[frame=single, linenos, fontsize=\footnotesize, breaklines]{sql}
      select max(a.salary) as secondHighest
      from employee as a, employee as b
      where a.salary < b.salary
    \end{minted}
    \vspace{-4mm}
     \captionsetup{font=small,name=Figure}
    \caption*{Q1: Cartesian product}
    \vspace{3mm}
  \end{minipage}
  % \vspace{2mm}
  \hfill
  \begin{minipage}{.45\textwidth}
    \begin{minted}[frame=single, linenos, fontsize=\footnotesize, breaklines]{sql}
      select max(salary) as secondHighest
      from employee
      where salary < (select max(salary) from employee)
    \end{minted}
    \vspace{-4mm}
     \captionsetup{font=small,name=Figure}
    \caption*{Q2: with subquery}
  \end{minipage}
  \vspace{-3mm}
  \captionsetup{font=small,name=Figure}
  \caption{Finding the second highest salary}
  \inv
  \label{fig:sql-comparison}
\end{figure}

\head{Exmaple 1: LeetCode \#176} The goal is to find the second highest salary from the Employee table. \query1 in Figure~\ref{fig:sql-comparison} is the human-written solution (submitted by LeetCode users) while \query2 emerges as a rewrite generated by LLMs. \query2 is approximately 2000x faster than \query1 when executed on a database  populated with one million rows. 
The optimization process involves a deep understanding of the query intent through reasoning: (1) identifying that the goal of \query1 is to find the second-highest salary, (2) realizing that there is an alternative way to achieve the same thing, i.e., first find the maximum salary using a subquery and then compare all salaries with this maximum salary to find the second maximum, and finally (3) recognizing that \query2 achieves performance improvement by reducing the number of comparisons required. No existing rewrite rule explicitly captures this transformation. As a result, this optimization cannot be achieved using traditional rule-based query rewriting techniques, highlighting the advantage of LLM-guided rewriting in discovering non-trivial, high-impact transformations.

\begin{listing}[ht]
  \tiny  % Reduce font size (use \footnotesize, \scriptsize, or even \tiny)
  \centering
  \begin{minipage}[t]{0.49\linewidth}  % First half
    \begin{tcolorbox}[colback=gray!5, colframe=gray!40, boxsep=2mm, left=6pt, right=5pt, top=0.5mm, bottom=0.5mm]
      \begin{minted}[
          frame=none,
          framesep=2mm,
          linenos,
          numbersep=2mm,
          escapeinside=||,
          baselinestretch=1.0, 
          fontsize=\tiny
      ]{sql}
with year_total as (
    select ..., 's' sale_type
    from  |\textcolor{blue}{$store\_sales$}|, ...
    ...
    union all
    select ..., 'w' sale_type
    from  |\textcolor{blue}{$web\_sales$}|, ...
    ...
)
select
    s_2.customer_id, ...
from
    year_total s_1,
    year_total s_2,
    year_total w_1,
    year_total w_2
where
    s_2.customer_id = s_1.customer_id
    and s_1.customer_id = w_2.customer_id
    and s_1.customer_id = w_1.customer_id
    and s_1.sale_type = 's'
    and s_1.dyear = 1998
    and w_1.sale_type = 'w'
    and w_1.dyear = 1998
    and s_2.sale_type = 's'
    and s_2.dyear = 1999
    and w_2.sale_type = 'w'
    and w_2.dyear = 1999
    and w_2.y_total / w_1.y_total > 
        s_2.y_total / s_1.y_total
      \end{minted}
    \end{tcolorbox}
    \vspace{-1mm}
    \caption*{(a) TPC-DS Q11 Original}  % Sub-caption inside the same listing
  \end{minipage}
  \hfill
  \begin{minipage}[t]{0.49\linewidth}  % Second half
    \begin{tcolorbox}[colback=gray!5, colframe=gray!40, boxsep=2mm, left=6pt, right=5pt, top=0.5mm, bottom=0.5mm]
      \begin{minted}[
          frame=none,
          framesep=2mm,
          linenos,
          numbersep=2mm,
          escapeinside=||,
          baselinestretch=1.0, 
          fontsize=\tiny
      ]{sql}
with store_sales_total as (
        select ...
        from |\textcolor{blue}{$store\_sales$}|, ...
        ...
    ),
    web_sales_total as (
        select ...
        from |\textcolor{blue}{$web\_sales$}|, ...
        ...
    )
select
    s_2.customer_id, ...
from
    store_sales_total s_1,
    store_sales_total s_2,
    web_sales_total w_1,
    web_sales_total w_2
where
    s_2.customer_id = s_1.customer_id
    and s_1.customer_id = w_2.customer_id
    and s_1.customer_id = w_1.customer_id
    and s_1.dyear = 1998
    and w_1.dyear = 1998
    and s_2.dyear = 1999
    and w_2.dyear = 1999
    and w_2.y_total / w_1.y_total > 
        s_2.y_total / s_1.y_total
      \end{minted}
    \end{tcolorbox}
    \vspace{-1mm}
    \caption*{(b) TPC-DS Q11 Rewritten}
  \end{minipage}
  \vspace{-1mm}
  \captionsetup{font=small,name=Listing}
  \caption{Left: Original TPC-DS Q11 query. Right: A rewrite of Q11 generated by LLMs. The rewrite is 9x faster, though no existing rewrite rules can achieve this transformation.}
  \label{listing:q11}
  \inv
\end{listing}
% \inv

\head{Example 2: TPC-DS Q11} In Listing~\ref{listing:q11} (left), a query adapted from TPC-DS Q11 identifies customers whose web sales growth from 1998 to 1999 outpaces their store sales growth over the same period. It begins by creating a Common Table Expression (CTE) \texttt{year\_total}, which aggregates annual store and web sales data for each customer (lines 2–4 and 6–8) via a \sqlunionallcolor operation (line 5). The query then filters this aggregated data by sales type and year to isolate the figures for 1998 and 1999 (lines 21–28). A join on \texttt{customer\_id} (lines 18–20) allows for a comparison of each customer's sales across different years and channels. In essence, this query uses \sqlunionallcolor to combine multiple data sources with a column to label each source, and then filtering on the label later to separate them. The combining and filtering cancel each other's effect and thus can be removed for better performance and readability.
 The revised query in Listing~\ref{listing:q11} (Right) makes two changes: (1) creates separate CTEs for store and web sales data, and (2) removes the sale type filters. This results in a 9x speedup, reducing execution time from 18 minutes to 2 minutes on PostgreSQL with TPC-DS scale factor 10. The performance boost is due not only to avoiding unnecessary computations but also to PostgreSQL's cardinality estimation errors with more predicates.

No existing rewrite rules address this anti-pattern, preventing rule-based query rewriting techniques from discovering this more efficient rewrite. However, a database expert reviewing the example would intuitively reason that the \sqlunionallcolor operation and the subsequent filtering cancel each other out. However, manually scrutinizing every query for new anti-patterns is not feasible. LLMs with advanced reasoning capabilities offer a promising alternative for automatically identifying and exploiting new rewrite opportunities.

\subsection{Large Language Models: Background}

Recent advancements in deep learning and the availability of large datasets have propelled LLMs into the spotlight. These models  exhibit advanced reasoning skills, allowing them to take on open-ended tasks and make logical inferences. 

\head{Prompt engineering}
The key to leveraging LLMs effectively lies in \emph{prompt engineering}\cite{googleprompt}. This involves crafting precise inputs to guide the model's output. By providing natural language instructions, users can guide the model towards the desired outcome. Through iterative refinement, prompt engineering allows for high-quality and relevant responses, giving better control over creativity and consistency. In query rewriting, we concatenate an input query $q$ with a prompt $p$, denoted as $p \texttt{||} q$, which encodes a comprehensive task description for rewriting. Figure~\ref{fig:basic_prompt} is an example of a basic prompt with only the query and a concise prompt,  which we refer to as \baseline.

\begin{comment}
\head{Zero-shot prompting vs few-shot prompting} The key to successfully using LLMs is to find the optimal prompt, commonly known as \emph{prompt engineering}\cite{googleprompt}. 
Prompt engineering is classified into two scenarios, depending on 
the number of examples provided in the prompt: zero-shot prompting\cite{gao2023text} and few-shot prompting~\cite{gu2023few}. In zero-shot prompting, no example is provided. Figure~\ref{fig:basic_prompt} shows a simple prompt which only includes the input query itself and a concise one-sentence rewriting instruction. Conversely, in few-shot prompting, a limited number of examples are provided to the LLM, allowing it to identify explicit or implicit patterns from the input examples and generate corresponding output. For instance, in the context of Text-to-SQL, an example in the few-shot prompt may include a database schema, a natural language query and the corresponding SQL queries as a solution.

\end{comment}

\head{Tokens and cost} Tokens are the units or building blocks that LLMs use to process language. For example, the string `` tokenization'' is decomposed as two tokens `` token'' and ``ization'', while a short common word like `` the'' is represented as a single token~\cite{openaitokens}. The cost of using LLMs is typically determined by the number of tokens in both the input and the output.

% \vspace{2mm}
\inv

\section{Lessons Learned from employing LLMs for query rewriting}
\label{sec:insight}

We conducted a comprehensive experiment to study the applicability of LLMs for  rewriting of a variety of queries  and improving their performance. Specifically, we carefully reviewed each rewrite suggested by LLMs and evaluated it for correctness and latency. We paid particular attention to discerning which problems could be effectively addressed and which could not, along with identifying conditions under which the latter could be transformed into the former.  Our study not only provided valuable insight int the capabilities of LLMs and the challenges associated with using it for query rewriting but also contributed to the development and refinement of \tool that is presented in \S\ref{sec:approach}.
Below we summarize the main lessons that have informed \tool's design decisions.

\head{Lesson 1: LLMs are effective on simple workloads but struggle with complex rewrites} While LLMs can identify effective rewrites for simpler queries using only basic prompts, their ability to recognize and apply meaningful transformations diminishes as query complexity increases.
Although it remains possible for LLMs to discover rewrite opportunities, the decreasing likelihood necessitates more iterations of LLM runs, resulting in significantly higher costs. In \S\ref{sec:motivation}, we demonstrated how Q11 in TPC-DS workload could be optimized into a more efficient form by eliminating the unnecessary \sqlunionallcolor. Similar inefficiencies exist in other TPC-DS queries, such as Q4 and Q74. Among them, Q4 stands out as the most complex: it has the highest token count (1,170 vs. 723 in Q11 and 591 in Q74), the greatest number of \sqlunionallcolor segments (3 vs. 2 for the others), and the most joins (5 vs. 3 for the others). Despite ten attempts using the prompt shown in Figure~\ref{fig:basic_prompt} with GPT-4, none of the rewrites for Q4 suggested removing the \sqlunionallcolor, a transformation that would have significantly reduced execution time. In contrast, 4 out of 10 attempts for Q11 and 6 out of 10 for Q74 moved in the correct rewriting direction. Note that a correct rewriting direction, such as the suggestion to eliminate unnecessary \sqlunionallcolor in Q74's rewrites, does not guarantee an error-free or equivalent rewrite to the original query. The inherent complexity of such queries makes it much more difficult for LLMs to generate fully correct rewrites, even when the high-level transformation logic is sound.

\head{Lesson 2: LLMs can optimize queries that they previously struggled to optimize when provided with suitable rewrite hints in the prompt}
As zero-shot prompting is not universally effective for query rewriting task,  it becomes essential to incorporate precise guidance within the prompts. Drawing inspiration from the few-shot prompting used in Text-to-SQL tasks~\cite{sun2306sql}, one approach is to provide LLMs with examples that include both the input query and its optimized version. Yet, an alternative strategy ---using concise, human-readable rewrite hints--- also holds promise.
For example, one may intuitively recognize that both queries can benefit from ``avoid the unnecessary \sqlunionallcolor''. In fact, integrating this insight as a hint in the prompt consistently leads LLMs to suggest removing \sqlunionallcolor across different attempts. This latter method offers two benefits over the few-shot prompting approach: (1) it requires fewer tokens, thereby reducing costs and accelerating the rewrite generation process, and (2) the hints are easily interpretable by humans, facilitating quicker debugging. However, due to the time constraints of human experts, manually analyzing each query and suggesting beneficial rewrite hints is impractical. Therefore, there is a need for an automated method to identify rewrite hints tailored for each query.

\begin{figure}[t]
\inv
  \centering
  \fbox{
    \begin{minipage}{0.8\linewidth}
        % This is a block of text inside a box in a figure in \LaTeX.
        % \textcolor{blue}{[input query]}\\
        \colorbox{gray!10}{[input query]}\\
      Rewrite this query to improve performance.
    \end{minipage}
  }
  \inv
  \captionsetup{font=small,name=Figure}
  \caption{The baseline approach for LLM-based query rewriting (\baseline)}
  \label{fig:basic_prompt}
 \inv
\end{figure}

\section{Our Approach}\label{sec:approach}

In this section, we describe our proposed algorithm, which incorporates the lessons presented in \S\ref{sec:insight}.

The iterative rewriting pipeline in \tool consists of three core components: (1) suggesting high-quality candidate rewrites, (2) iteratively correcting these rewrites for equivalence, and (3) evaluating their effectiveness. This pipeline closely integrates with \tool's \textbf{Natural Language Rewrite Rule (NLR2) Repository}, enabling improved knowledge transfer and more fine-grained control over the query rewriting process.

% Additionally, it uses a natural language rewrite rule (NLR2) repository to facilitate better knowledge transfer and a more fine-grained control over the rewriting process.

\tool's top-level algorithm is presented in Algorithm~\ref{algo:tool}. Initially, the algorithm takes a set of queries $\mathcal{Q}$ as input for optimization. NLR2 Repository $\mathcal{R}$ may either be initialized as empty or pre-populated with an offline-constructed repository $\mathcal{R}_{pre}$. At each iteration, \tool explores the NLR2 repository to identify the most relevant NLR2s for the target query, integrating them into the rewriting prompt. If no suitable NLR2s are found,\tool defaults to using a basic prompt. The LLM suggests candidate rewrites, accompanied by explanations describing the applied NLR2s and the conditions under which they are applicable. Through iterative correction and feedback mechanisms, \tool progressively converges towards generating more accurate rewrites, thereby largely mitigating the issue of inequivalence. Semantic correction feedback is obtained from language models, while syntax correction relies on feedback from the database. Subsequently, our tool evaluates rewrites for equivalence and performance. Equivalence is assessed using an off-the-shelf tester or verifier, whereas performance is evaluated either by actual execution or by estimation through a database cost model. It also updates the selected rules and their benefits in the NLR2 repository. With each iteration, a larger set of optimized queries and a more comprehensive collection of NLR2s with more precise benefit estimates are obtained. Eventually, once the set of optimized queries stabilizes (or the budget is exhausted), the rewritten queries are presented to the user for final inspection.

\begin{algorithm}
  \caption{\tool top-level Algorithm}\label{algo:tool}
  \SetAlgoNlRelativeSize{-1}
  \SetAlgoNlRelativeSize{-2}
  \SetAlgoNlRelativeSize{-1}
  
  \KwIn{$\mathcal{Q}$: queries, $\mathcal{L}$: LLM, $\mathcal{D}$: database, $\mathcal{T}$: tester, $B$:budget,
   $\theta$: user-specified minimum desirable speedup}
  \KwOut{$\mathit{Res} = \{\langle q, q', e\rangle | \forall q \in \mathcal{Q}_{opt} \subseteq \mathcal{Q}$, where $\mathcal{Q}_{opt}$ is the optimizable subset of $\mathcal{Q} \}$: Optimized queries \& associated NLR2s}
  
  \SetAlgoNlRelativeSize{-2}
  \SetAlgoNlRelativeSize{-1}
  \SetAlgoNlRelativeSize{-1}
  \SetAlgoNlRelativeSize{-1}
  
  % \tcp{\textcolor{commentgray}{Extract context from queries}}
  $\mathit{Res} \leftarrow \{\}$\;
  $\text{NLR2 Repostitory } \mathcal{R} \leftarrow \{\}  \text{ or pre-built repository } \mathcal{R}_{pre}$\;
  \While{$\mathcal{Q}$ is not empty}{
    \For{each query $q$ in $\mathcal{Q}$}{
        % $SR \leftarrow$ Select-Rules ($q, \mathcal{R}$)\; 
        $\tilde{q}, e \leftarrow$ Suggest-and-explain ($q, \mathcal{L}, \mathcal{R}$)\;
        $q' \leftarrow$ Correct-for-equivalence ($q, \tilde{q}, \mathcal{L}, \mathcal{D}$)\;
        % \If{$equiv$ is true}{
            $equiv, speedup \leftarrow$ Evaluate-rewrite ($q, q', \mathcal{D}, \mathcal{T}$)\;
            \If{$equiv$ is true}{
            $\mathcal{R} \leftarrow$ Update-NLR2-repo ($\mathcal{R}, e, speedup$)\;
            \If{$speedup > \theta$}{
                % $\mathcal{Q}$.remove ($q$)\;
                $\mathit{Res}$.add ( $\langle q, q', e\rangle$ )\;
                % Move $q$ from $Q$ to $Q'$\;
            }
            % }
        }
    }
    \If{$\mathit{Res}$ does not change \emph{\textbf{or}} budget $B$ is exhausted}{
        \KwRet{$\mathit{Res}$}\;
    }
    Remove queries in $\mathit{Res}$ from $\mathcal{Q}$\;
}
  \SetAlgoNlRelativeSize{-1}
\end{algorithm}
\inv

\begin{figure}[t]
\inv
  \centering
  \fbox{
    \begin{minipage}{0.8\linewidth}
        % This is a block of text inside a box in a figure in \LaTeX.
      % \textcolor{blue}{[input query]}\\
      \colorbox{gray!10}{[Input query]}\\
      Rewrite this query to improve performance. Only use this rule when rewriting: \colorbox{gray!10}{[a selected NLR2]}

      % \textcolor{blue}{[a selected NLR2]}
    \end{minipage}
  }
  \inv
  \captionsetup{font=small,name=Figure}
  \caption{A prompt incorporating a selected NLR2.}
  \label{fig:prompt_with_nlr2}
  \inv
\end{figure}
\inv

% For each input query, we employ an LLM to generate candidate rewrites. Simultaneously, we log the associated NLR2s and conditions for each rule's application. If a rewrite proves effective, it is uploaded to the rule repository, where its benefit is updated. Different from the zero-shot setting in base case, we actively search the Rule Repository for the most promising NLR2s to include as hints in the prompt, thereby enhancing the guidance for candidate generation. Subsequently, by iteratively refining the candidates and incorporating feedback mechanisms, \tool can progressively converge towards generating more accurate rewrites, thereby largely alleviate the issue of inequivalence. Semantic refinement feedback is sourced from LLMs, while syntactic refinement relies on the database for feedback. In light of limited time budget of DBAs, we prioritize the most promising rewrites for their review based on the benefit of applied NLR2s. 

\subsection{\jie{Natural Language Rewrite Rules (NLR2s)}}
\label{sec:approach:nlr2}

Since basic prompts do not consistently yield effective results, especially for highly complex queries, it is essential to incorporate proper hints into the prompt to guide the LLM. In \tool, hints are represented as  \textbf{Natural Language Rewrite Rules (NLR2s)}. The corresponding prompt format is presented in Figure~\ref{fig:prompt_with_nlr2}.

\jie{
NLR2s encapsulate beneficial rewrites using natural language, differing fundamentally from traditional pattern-matching rewrite rules, which define a pattern-matching transformation and constraints specifying when that transformation is safe. Instead, NLR2s offer greater flexibility and expressiveness, as they are not restricted to matching fixed patterns but can capture nuanced, contextual insights into when and how a query should be rewritten.}

% \vspace{-10pt}

\head{\jie{Examples of NLR2s} } 
Table~\ref{tab:nlr2-speedups} highlights three example NLR2s that yield significant speedups for Q11 (Listing~\ref{listing:q11}). These transformations are mutually exclusive, each producing a distinct query structure and logical flow. Notably, the NLR2 addressing \sqlunionallcolor redundancy in Section~\ref{sec:motivation} corresponds to  $r_3$. The remaining two transformations leverage different early data filtering ideas, reducing the size of intermediate results and enhancing overall efficiency. None of these transformations can be captured by existing pattern‐matching rules.

% \inv
 We meticulously curate a set of NLR2s in the NLR2 repository. These NLR2s fulfill a dual role: they act as both guiding hints for LLMs and explanations for users. When incorporated into the prompts for LLMs, these rules provide appropriate guidance for the rewriting direction. Meanwhile, presenting these NLR2s to users can help them to better understand the rationale behind the proposed modifications.  \jie{With the prompt in Figure~\ref{fig:prompt_with_nlr2}, LLMs prioritize implementing the provided NLR2s over relying on own creativity. The quality and relevance of these NLR2s directly impacts the effectiveness of the rewrites. However, } 
manually analyzing each query and formulating effective NLR2s is impractical. Therefore, there is a need for an automated pipeline to identify promising NLR2s.

\begin{table}[tbp]
\inv
    \centering
    \small
    \renewcommand{\arraystretch}{1.2} % Increase row height
    \setlength{\tabcolsep}{6pt} % Increase column padding
    \begin{tabular}{|l|l|l|}
        \hline
         & \textbf{NLR2} \textbf{as Rewrite/Transformation hints} & \textbf{Speedup} \\
        \hline
        r1 & Split aggregated computations by filtering  & 24.5x \\
           & on the specific period in separate subqueries & \\
        \hline
        r2 & Combine multiple self-joins by pivoting  & 11.7x \\
           & aggregated data into columns using CASE & \\
        \hline
        r3 & Eliminate union operations by computing  & 9x \\
           & aggregates in dedicated subqueries per data source & \\
        \hline
    \end{tabular}
    \captionsetup{font=small,name=Table}
    \caption{Natural Language Rewrite Rules (NLR2s) that benefits TPC-DS Q11 query with their corresponding speedups.}
    \inv
    \label{tab:nlr2-speedups}
\end{table}

% \subsubsection{\textbf{Collecting NLR2s}}

\head{Collecting NLR2s} 
% In order to maximize the efficacy of LLMs, the NLR2 we provide as the hint should be easily understood by the LLM. We adopt a simple but effective method of collecting useful NLR2s: let the LLM summarize 
 Our approach is simple yet effective: we prompt the LLM to generate candidate rewrites while simultaneously summarizing the underlying NLR2s. If a rewrite proves effective in subsequent evaluations, we consider its associated NLR2 valuable and store it—along with the corresponding query features and observed speedup—in the NLR2 Repository for future use. We will discuss NLR2 evaluation and query feature extraction  later. To elicit rewrite rules, we append the instruction ``Describe the rewrite rules you are using'' to the basic prompt template in Figure~\ref{fig:basic_prompt}. Moreover, if extracted NLR2s contain concrete column or table names, they may degrade the quality of future rewrites. To mitigate this, we explicitly instruct the LLM to exclude query-specific details by adding ``do not include any specific query details in the rules, e.g., table names, column names'' to the prompt, ensuring the NLR2s remain as general and reusable as possible.

\cut{
If candidate rewrites prove to be effective, corresponding NLR2s are very likely to provide significant value, and we tend to apply these rules to similar queries for better performance. However, if the NLR2s extracted from previous queries contains different column names or table names compared to the incoming queries, they may confuse the LLM when used as hints, potentially lowering the quality of the rewrites. Therefore, NLR2s should be relatively general, not tied to a specific query but rather focusing on the transformation itself. To mitigate this issue, we explicitly instruct the LLM to exclude query-specific information in the prompt. Furthermore, we introduce an additional round of prompting to prompt the LLM to specify the conditions for applying the rule, providing us with more insights into the rule and aiding in a more thorough analysis of the collected rules in subsequent processing steps.

Without any input from human experts, we are able to collect candidate rewrites and associated NLR2s for each input query through the utilization of the LLM. Before using them as rewrite hints, however, we need to gain a better understanding of the potential benefits each rule can offer, which will be addressed in the next section.}

\head{Measuring the Benefit of NLR2s} In traditional query optimization, multiple rewrite rules can be applied on the same query as long as the changes they make do not conflict with each other. Similarly, we may collect multiple NLR2s associated with one candidate rewrite $q'$. In our experiments on the TPC-DS benchmark, we observe that each rewrite typically corresponds to 3–6 NLR2s. However, both generating rewrites with LLMs and executing complex queries can be computationally expensive. It is therefore essential to prioritize NLR2s that yield meaningful performance gains and avoid wasting resources on ineffective ones.

We observe two key challenges that hinder efficient and accurate estimation of an NLR2’s benefit.

% However, both generating rewrites with LLMs and executing complex queries can be computationally expensive. It is therefore essential to prioritize NLR2s that yield meaningful performance gains and avoid wasting resources on ineffective ones.
% For a query in TPC-DS benchmark, this number is in the range of 3 and 6.
% Utilizing LLMs for complex tasks like query rewriting and running complex queries can both  be costly. Therefore, it is crucial to avoid spending resources on NLR2s that offer little to no improvement.
% We observe two problems that make this benefit estimation process hard and inefficient.
% We need to evaluate the benefits of individual NLR2s and identify the most effective ones for future rewrites.

\head{Observation 1} A single rewrite $q'$ is often associated with multiple NLR2s, each affecting query performance differently. This makes it challenging to pinpoint the precise contribution of each rule to any observed performance improvement. Assuming that all NLR2s associated with the same rewrite are equally effective is unrealistic and could undermine the  effectiveness of our prompting strategy.

\head{Observation 2} Two NLR2s can be functionally identical despite having different descriptions. For example, the NLR2s in Table~\ref{tab:similar_rules} all target the same performance issue caused by correlated subqueries. Failing to recognize such equivalence can lead to redundant evaluations, wasting both resources and time.

\begin{table}[tbp]
\inv
    \renewcommand{\arraystretch}{1.25} % Adjust the value to increase or decrease the row height
    \centering
    \small
    \begin{tabular}{|c|p{7cm}|}
        \hline
        $r_a$ & Replace the correlated subquery with a precomputed aggregate in a separate CTE \\
        \hline
        $r_b$ & Use common table expressions (CTEs) to precompute aggregates \\
        \hline
        $r_c$ & Replace a correlated subquery that computes an aggregate with an inline view (or common table expression) that pre-aggregates data \\
        \hline
        $r_d$ & Replace a correlated subquery with a CTE for aggregated values \\
        % \hline
        % 5 & Use explicit join conditions. \\
        % \hline
        % 6 & Move conditions from WHERE clause to ON clause in JOINs. \\
        \hline
    \end{tabular}
    \captionsetup{font=small,name=Table}
    \caption{Examples of a set of  NLR2s that are semantically equivalent to each other}
    \label{tab:similar_rules}
    \inv
\end{table}

% miscalculating the utility of specific NLR2s. Consider a scenario where applying Rule A from Table~\ref{tab:similar_rules} to a query results in a 1000x speedup, while applying Rule B to another query yields no noticeable difference. One might mistakenly conclude that Rule A is significantly more valuable than Rule B. However, these two rules should be regarded as one. Recognizing this helps us better understand the potential impact this NLR2 has in different contexts. 

We introduce two techniques to address the issues mentioned above: \textbf{\textit{plan-based dominant NLR2 identification}} and \textbf{\textit{NLR2 grouping via LLMs}}.

\head{Plan-based dominant NLR2 identification} 
When the LLM produces a faster rewrite $q'$ for a given input query $q$, it also returns a set of associated NLR2s, denoted as $\{r_1, r_2, \ldots, r_k\}$. Instead of treating all rules equally, we aim to identify a single dominant NLR2, denoted $r^*$, that primarily accounts for the observed performance gain. This allows us to attribute the speedup to the most impactful transformation, rather than distributing credit uniformly across all contributing rules. To identify the dominant NLR2 $r^*$, we prompt the LLM with each NLR2 individually on the same input query $q$, producing candidate rewrites ${q_{r_1}, q_{r_2}, \ldots, q_{r_k}}$. For each $q_r$, we obtain its query plan $P(q_r)$ via \texttt{EXPLAIN} (which reports the plan without executing the query) and compare it to the plan of $q'$, denoted $P(q')$. \revone{We first check for an exact cost match: if there exists a rule $r$ such that the optimizer’s cost estimate for $q_r$ equals that of $q'$ (i.e., $\mathcal{C}(q_r)=\mathcal{C}(q')$), we set $r ^* = r$, since identical costs typically indicate identical plans. Otherwise, we choose the NLR2 whose plan is most similar to $P(q')$:}
    \[
    \revoneeq{
    {r^*=\arg \min _{r \in \mathcal{R}} d_{\text {plan}}\left(P\left(q_r\right), P\left(q^{\prime}\right)\right)
    }}
\]
\noindent\revone{where plan similarity is measured by embedding each \texttt{EXPLAIN} plan using \texttt{bert-base-uncased} and taking the $\ell_2$ distance between the embeddings: $d_{\text {plan }}\left(P_1, P_2\right)=\left\|\phi\left(P_1\right)-\phi\left(P_2\right)\right\|_2$.} We then attribute the observed speedup of $q'$ over $q$ to $r^*$ and treat that improvement as the estimated benefit of the dominant rule.

\revone{The identification process relies on exactly one piece of execution, i.e., the improved rewrite $q'$, which serves as the anchor. We do not execute any additional candidate rewrites ${q_{r_1}, q_{r_2}, \ldots, q_{r_k}}$; instead, we rely on inexpensive plan and cost comparisons.} This assumption of a single dominant NLR2 rewrite simplifies the real-world scenario, where interactions between rules can occur. In practice, a single rule may yield little benefit in isolation but unlock further optimizations when combined with others. However, exhaustively evaluating all rule combinations is computationally prohibitive. Our method offers a scalable approximation.

% This approach leverages existing query execution history to efficiently evaluate candidate rewrites ${q_{r_1}, q_{r_2}, \ldots, q_{r_k}}$ without executing each against the database. While this method simplifies the real-world scenario, where interactions between rules may exist, it offers a practical tradeoff. In some cases, a single rule may yield little benefit in isolation but unlock further optimizations when combined with others. Exhaustively evaluating all rule combinations, however, is computationally infeasible. Thus, identifying the dominant NLR2 through independent evaluation provides a scalable and effective approximation.

\begin{figure}[t]
\sinv
  \centering
  \small
  \fbox{
    \begin{minipage}{0.9\linewidth}
      \textbf{Input:}\\
      % Replace implicit joins with explicit joins \textcolor{blue}{$\leftarrow$the incoming NLR2}  \\
      Replace implicit joins with explicit joins \textcolor{blue}{$\leftarrow$the incoming NLR2}  \\
    Please select the rewrite rule that is strictly the same as the above rule and give your explanation (just give one answer). If not, please select the first item “Unseen rule”.\\
    Options:\\
1. Unseen rule\\
2. Replace subqueries with join\\
3. Use explicit join syntax instead of comma-separated tables in the from clause\\
4. Use a CTE to calculate the average

\textbf{Answer:}\\
Use explicit join syntax instead of comma-separated tables in the from clause \textcolor{blue}{$\leftarrow$LLM's choice after reasoning}

Explanation: Both the original rule and this rewritten rule are about replacing implicit joins with explicit joins. \textbf{Implicit joins are indicated by comma-separated tables in the FROM clause and join conditions in the WHERE clause}. ...
      
    \end{minipage}
  }
  \inv
  \captionsetup{font=small,name=Figure}
  \caption{The prompt to predict the group to which the incoming NLR2 belongs}
  \label{fig:clustering}
  \inv
\end{figure}

\head{NLR2 grouping via LLMs} 
While plan-based dominant NLR2 identification efficiently leverages query execution history to avoid running expensive queries on real data, repeatedly invoking the LLM on $q$ with different NLR2s as hints incurs a noticeable amount of API cost. As noted in \textbf{Observation 2}, the set of NLR2s extracted from a single query $q$ are often not sufficiently distinctive. Supplying semantically equivalent hints to the LLM yields little additional benefit while unnecessarily increasing the costs. To address this, every time a new NLR2 $r_{\text{new}}$ is discovered for query $q$, we leverage LLMs to determine its semantic equivalence to existing NLR2s associated with $q$. If it is deemed equivalent to a previously identified group, we avoid invoking the LLM again for that rule; otherwise, a new group is created for $r_{\text{new}}$, and we proceed to run the LLM rewriting pipeline to generate the corresponding rewrite $q_{r_{\text{new}}}$ and obtain its query plan. This allows us to evaluate whether $r_{\text{new}}$ qualifies as the dominant NLR2 for query $q$.

% To evaluate the effectiveness of a specific NLR2 on a given query, we include it in the prompt, then run the query through the iterative correction process by LLMs, and finally measure its latency via actual execution. Although this pipeline is both time- and cost-intensive, it enables us to discover rewrite opportunities that can later be converted into database rules if they offer significant benefits. However, invoking LLMs on queries following the same template may yield a collection of NLR2s that are not sufficiently distinctive. Repeatedly feeding semantically equivalent NLR2s to the same query does not provide any additional benefit but only increases the overall computational and cloud database costs.

% , thereby identifying the appropriate group for it. If the LLM finds no semantically equivalent candidate, a new group is created for the NLR2. 

 Figure~\ref{fig:clustering} illustrates an example prompt used for this grouping process. Replacing implicit joins with explicit joins hardly yields any noticeable performance gap, yet such transformations are often included among the collected NLR2s. Grouping NLR2s not only prevents redundancy in our repository but also simplifies maintenance and ensures that our efforts are concentrated on rewrite strategies that deliver substantial performance gains.

\subsection{Suggesting Candidate Rewrites}
\label{sec:approach:suggest}

In the previous section, we discussed how to construct a high-quality NLR2 Repository over time. The primary goal of this repository is to enable effective knowledge transfer, particularly for optimizing complex queries that vanilla LLM-based approaches fail to handle. However, NLR2s that yield substantial speedups for certain queries may degrade performance when applied to others. A key challenge, therefore, is to identify queries that share similar performance bottlenecks, so that we can selectively incorporate the most relevant NLR2s from the repository into the prompt for effective and \revone{equivalent} rewriting.

\head{Query performance bottleneck analysis}
Since we focus on optimizing recurring queries with readily available query plans, we aim to extract as much insight as possible from these plans. To this end, we leverage LLMs to inspect query plans and identify potential performance bottlenecks. Prior to analysis, we preprocess each plan to remove irrelevant fields, thereby reducing the number of tokens and minimizing distractions from non-essential details. For example, in PostgreSQL, eliminating over 20 unnecessary fields can reduce token count by up to 43\%. We then provide the LLM with the prompt consisting of three components: the original query $q$, the trimmed query plan, and a concise instruction: “Summarize the most critical performance bottleneck based on the query and its plan.”
It is important to note that the query plans used for bottleneck analysis are retrieved from execution history and include actual runtime metrics (e.g., actual time, rows, loops) for each node. In contrast, the plans used for identifying dominant NLR2s only include the planner’s chosen execution plan and its estimated costs, without any observed runtime statistics.

\head{Similarity search}
Using the bottleneck summary generated by the LLM, we perform a two-stage similarity search to identify relevant past queries. (1) we encode the bottleneck summary into a feature embedding and retrieve the top-$k$ most similar embeddings from previously analyzed queries. (2) we apply a refinement step inspired by the method illustrated in Figure~\ref{fig:clustering}: we prompt the LLM with a multiple-choice question to identify which of the retrieved candidates shares the most similar performance bottleneck with $q$.  If the LLM determines that none of the candidates are a good match, the system falls back to using the basic prompt without incorporating any prior NLR2. Based on the selected candidate, we then retrieve and apply its most beneficial associated NLR2.

\cut{Since simple zero-shot prompts do not consistently yield effective results, especially for highly complex queries, it is essential to incorporate proper hints into the prompt to guide the LLM. In \tool, hints are represented as a set of  NLR2s. }

\cut{\textbf{Natural Language Rewrite Rules (NLR2s)} differ fundamentally from traditional pattern-matching rewrite rules. A traditional rewrite rule is comprised of two key components: (1) a transformation that maps a source query pattern to a corresponding destination query pattern, and (2) a set of constraints under which the rule can be safely applied. These pattern-matching rules are typically described using a domain-specific language, which can be challenging for both humans and LLMs to understand. Directly integrating rules presented in the internal language as rewrite hints therefore do not provide the necessary guidance for the LLM. Instead, our NLR2s briefly summarize beneficial rewrites or transformations in natural language. Compared to pattern-matching rewrite rules, NLR2s offer enhanced flexibility and expressiveness. Below are some example NLR2s used in \tool:}

% \begin{itemize}
% \item Pre-calculate aggregates in subqueries
% \item Remove unnecessary UNION ALL operation
% \item Avoid using arithmetic operations in WHERE clause
% \item Replace implicit JOINs with explicit JOINs
% \item ...
% \end{itemize}

 \cut{Figure~\ref{fig:hints} shows an example prompt that is used to suggest candidate queries.}
  
  % as hints for LLMs to consider. These rules are human-readable, serving as explanations for the rewriting process alongside the rewrite candidates. Therefore, NLR2s are both hints and explanations. When presented as rewrite hints in the prompt to LLMs, these rules provide appropriate guidance. When presented as explanations alongside the rewrite candidates, users gain a clearer understanding of the transformations made to the input queries.  

\cut{Since manual inspection is often not feasible due to the limited time budget of human experts, we must devise a method to learn from the past instances, and automatically generate appropriate hints with minimal human intervention.}

% Zero-shot prompts are effective for a large set of queries but may struggle with highly complex queries. To enhance performance, additional guidance should be integrated into the prompt to assist LLM in generating more promising rewrites. However, relying solely on human experts for guidance is impractical. Instead, we aim to leverage an automated sources -- LLM, to provide the necessary guidance. More specifically, we need to automatically learn useful NLR2s from the instances LLMs can successfully solve and see if they can be utilized to optimize queries that LLM previously failed to solve. Figure~\ref{fig:hints} shows an example prompt that is used to suggest candidate queries. 

% \subsubsection{\textbf{Collecting NLR2s.}}

% instruct the LLM to describe the rewrite rules involved in rewriting the input query to the rewrite it presents

\begin{figure}[h]
\sinv
  \centering
  \small
  \fbox{
    \begin{minipage}{0.9\linewidth}
        % This is a block of text inside a box in a figure in \LaTeX.
        \textbf{Input:}\\
      q1:\colorbox{gray!10}{[Input query]}\\
      q2:\colorbox{gray!10}{[Candidate rewrite]}\\
      q1 is the original query, q2 is the rewritten query of q1.\\For q1, break it down step by step and then describe what it does in one sentence. Do the same for q2.\\
      Give an example, using tables, to show that these two queries are not equivalent if there's any such case. Otherwise, just say they are equivalent.\\
      \textbf{Answer:}\\
      Not equivalent. \\
      \colorbox{gray!10}{[Breakdown and analysis with \textbf{counterexamples}]} \\
      % $\{$Breakdown and analysis$\}$\\
      % $\{$A counterexample$\}$\\
      \textbf{Input:}\\
      Based on your analysis, which part of q2 should be modified so that it becomes equivalent to q1? Show the modified version of q2.\\
      \textbf{Answer:}\\
      % \textcolor{blue}{[Revised candidate rewrite]} 
      \colorbox{gray!10}{[Revised candidate rewrite]}

    \end{minipage}
  }
  \inv
  \captionsetup{font=small,name=Figure}
  \caption{The prompt for semantic correction}
  \label{fig:semantic_correction}
  \inv
\end{figure}

% for example, LLM would not apply ``remove unnecessary UNION ALL'' to a input query that does not contain UNION ALL

\subsection{Correcting Candidate Rewrites}
\label{sec:approach:correct}

Unlike traditional rewrite rules, which are verified for correctness, LLM-generated rewrites lack guaranteed accuracy, even with appropriate rewrite hints. While many tools can assess query equivalence with varying degrees of capability and efficiency, our requirements extend beyond this. Our goal is not only to verify the equivalence of the rewrite to the input query but also to correct the rewrite until it matches the original query. Otherwise, it is inefficient and a waste of computing resources to repeatedly prompt the LLM in hopes of producing a correct rewrite.

Errors generally fall into two categories: 1) Executable with incorrect outcomes, such as a mismatch where the input query aims to find the second-highest salary, whereas the candidate rewrite seeks the highest salary—this type of error is termed a semantic error. and 2) Unexecutable due to various issues, such as incorrect column names or table names—this type is referred to as a syntax error. These two categories of errors are relatively independent of each other. Inspired by these observations, we develop a two-stage counterexample-guided correction approach. During the semantic correction stage, we focus on ensuring that the candidate rewrite achieves the same intent as the input query, while temporarily setting aside syntax errors. In the syntax correction stage, we address all issues impacting query execution.

\head{Semantic Correction} Given the input query as the gold standard, we iteratively refine each candidate rewrite to ensure convergence toward semantic equivalence. 
In each iteration, the LLM first summarizes the intent and logic of both the original and candidate queries to understand their respective semantics. It then attempts to identify a counterexample that highlights a divergence in their results, leveraging the extracted summaries and  reasoning. Based on this comparison, the LLM determines whether the candidate rewrite is equivalent to the input query. If deemed ``not equivalent'', the LLM generates a revised version of the candidate query, informed by the identified discrepancies. This revised candidate is evaluated in the next iteration. The process continues until the LLM deems the candidate query equivalent to the input.
Figure~\ref{fig:semantic_correction} illustrates an example prompt used in the semantic correction phase.

% We adopt the idea of Chain-of-Thought prompting in our design: break down the problem-solving process into smaller, more manageable parts, making it easier for the model to understand and process the query. 

% 4 of 10 attempts for Q11 and 6 of 10 for Q74 moved towards a correct rewriting direction.

\cut{Using TPC-DS Q11 from Section 2 as an illustration, 4 of 10 rewriting attempts by GPT-4 for Q11 correctly identify the need to eliminate redundant \sqlunionallcolor operations and decompose the CTE into smaller ones. However, none of the rewrites are equivalent and error-free. For intricate queries like Q11, it is particularly challenging for the LLM to match the semantics of the input query precisely while avoiding syntax errors. The initial attempt at rewriting Q11 often results in fewer joins than the correct version, overlooking the essential role of these joins in facilitating the comparison of sales data for the same customer across different years and types of sales. Achieving a good rewrite of Q11 without semantic errors typically requires 1-3 iterations, even then, minor syntax errors  remained, which are addressed during the syntax correction phase.}

\head{Syntax Correction} 
% Syntax correction adopts a philosophy similar to semantic correction, with a notable distinction: syntax correction depends on feedback from executing the \texttt{EXPLAIN} command on the candidate rewrite to guide its refinement process. In contrast, semantic correction relies on counterexamples identified by the LLM through logical reasoning. Typical errors addressed include incorrect column or table names and inconsistent table aliases. The use of EXPLAIN, which does not execute the query, minimizes overhead, making it an efficient method for decide whether the rewrite is executable or not.
Syntax correction is also an iterative process, similar to semantic correction, but with a key distinction: syntax correction depends on feedback from executing the \texttt{EXPLAIN} command on the candidate rewrite to guide its refinement process, while semantic correction relies on counterexamples identified by the LLM through logical reasoning. At each iteration, we prompt the LLM with the original query $q$, the candidate rewrite $q'$, and the error message returned by the \texttt{EXPLAIN} command (if any). Typical errors addressed include incorrect column or table names and inconsistent table aliases. The use of EXPLAIN, which does not execute the query, minimizes overhead, making it an efficient method for decide whether the rewrite is executable or not.

% refine query based on the feedback returned by ``EXPLAIN'' command

\subsection{Evaluating Candidate Rewrites}
\label{sec:approach:evaluate}

Our ultimate goal is to identify equivalent queries that are \emph{faster}, \revthree{without introducing regressions}. To this end, we subject each candidate rewrite to \revthree{two automated gates: a correctness gate and a performance gate.}

\head{\revthree{Correctness gate}} We employ a combination of off-the-shelf verifiers and testers, selected based on the available time budget and the query complexity. Specifically, we use SQLSolver~\cite{ding2023proving}, a state-of-the-art SQL verifier, to assess the correctness of generated rewrites. If SQLSolver returns ``UNKNOWN'', we fall back to a tester from SlabCity~\cite{dong2023slabcity}, which extracts execution hints from the input query and applies syntactic analysis to guide input generation. Counterexamples generated during the semantic correction phase can also be incorporated into the tester to enhance coverage. \revthree{If neither the verifier nor the tester can certify equivalence, we abstain and discard the rewrite.}

\head{\revthree{Performance gate}} Only candidates that pass the correctness gate proceed to performance evaluation. \revthree{We run an offline shadow execution of  the candidate rewrite against the target database. A rewrite is accepted only if it satisfies a ``never-worse-off'' policy; otherwise, the rewrite is automatically rejected and the system falls back to the original query. We also pre-filter with a static cost proxy to avoid expensive trials.} This procedure is particularly justified for workloads with recurring queries of fixed periodicity. 

\revthree{In the exploratory setting, an optional manual review step can be used to expand coverage, but this is not required for the deployment.}

\cut{Compared to the simple zero-shot prompt in Figure. \ref{fig:basic_prompt}, our prompt for suggesting candidate queries incorporates two changes. The first change is to include a list of NLR2s as rewrite hints. We defer the details of how to select the most relevant NLR2s to later sections. However, it is not possible to select and apply the appropriate rules before we collect them, which is precisely the problem the second change aims to address.}
\section{Evaluation}
\label{sec:expr}

The evaluation aims to answer the following questions:
\begin{enumerate}[leftmargin=0.5cm]
       \item  How does \tool compare to a straighforward application of the out-of-the-box LLM? In other words, do our   NLR2s and counterexample-based correction algorithm considerably boost the LLM's performance? (\S\ref{sec:expr:llmbaseline})

\item How does \tool compare to state-of-the-art rule-based query rewriting approaches? Specifically, which approach can optimize more queries and which approach yields more speedup for the rewritten queries? (\S\ref{sec:expr:traditionalbaseline})

\item  How much does each of \tool's design contribute to its overall success (ablation study)?   (\S\ref{sec:expr:ablation})

\item  What is the time and monetary cost of \tool?  (\S\ref{sec:expr:llmcost})

% \item How does the choice of the LLM with various capabilities influence the effectiveness of \tool's rewrites? (\S\ref{sec:expr:llmchoice})
\end{enumerate}

\subsection{Experimental Setup}
\label{sec:expr:setup}

\head{Testbed} Our experiments are all conducted on Google Cloud Engine Machine n2-highmem-2 instances, with with 16GB RAM and 2 vCPUs. We use PostgreSQL 14.17. To evaluate query latency accurately, we execute each query 3 times and take their average as the final latency value. \revthree{We do not create any indexes. Instead, we run \texttt{VACUUM ANALYZE} on all tables before benchmarking, which collects statistics for cost-based plan selection.}

\cut{Before each run, we restart the PostgreSQL service and clear the database cache.} \cut{revise later, we have an off-shelf SQL verifier that works for some TPC-DS queries and most JOB queries Additionally, we use an off-the-shelf tester from~\cite{dong2023slabcity} to check query equivalence followed by manual inspection.}

% we generate a carefully designed dataset of small size to verify the correctness of our query rewrites.

\head{Choice of LLM} We designate OPENAI o3-mini as the default LLM, due to its cost efficiency and robust reasoning capabilities. We also assess \tool's generalizability across other LLMs with varying capabilities, such as GPT-4o and GPT-4-mini.

\head{Workloads} \revtwo{Our experiments include two widely used analytical benchmarks, TPC-DS~\cite{tpcds} and the Join Order Benchmark (JOB)~\cite{leis2015good}, as well as additional query sets generated by SQLStorm~\cite{schmidt2025sqlstorm} over the same TPC-DS and JOB schemas. SQLStorm uses LLMs to synthesize thousands of executable, high-complexity SQL queries and then filters them using actual database feedback. These SQLStorm queries were released in June 2025—after the pretraining cutoff of the LLM we use (October 2023)—and thus were not available during pretraining, allowing us to assess how well \tool generalizes to previously unseen queries on familiar schemas.}

% We evaluate \tool on multiple workloads to test its robustness across schema and query complexity. 

% To verify the effectiveness of \tool across different scenarios, we conduct experiments on two workloads: TPC-DS and JOB. Besides these two standard benchmarks, we also include queries generated by SQLStorm under the same public schema, which is unknown to the LLMs, so that we can test how our methods performs

\begin{enumerate}[leftmargin=0.5cm]

\item \textbf{TPC-DS.}
TPC-DS is an industry-standard decision support benchmark. While it does not perfectly capture the recurring, user-specific workloads seen in production, it remains a strong proxy for realistic analytical queries and is widely used to compare query optimization techniques. We generated a 10\,GB TPC-DS instance and used all 99 benchmark queries.

\item \textbf{Join Order Benchmark (JOB).}
JOB is a canonical benchmark derived from the IMDB dataset and is heavily used to study join ordering and cost-based optimization. Its queries span diverse levels of complexity and involve multi-way joins, making it a good stress test for optimizer behavior.

\item \textbf{\revtwo{TPC-DS (SQLStorm).}}
\revtwo{SQLStorm generated tens of thousands of query variants over the TPC-DS schema. Running LLM-based methods on all of them would be prohibitively expensive due to LLM inference cost. We therefore select the 100 slowest SQLStorm-generated queries as measured by runtime under \texttt{olapbench}~\cite{OLAPBench}. According to SQLStorm's own analysis, these queries fall in the medium-to-high complexity range.}

\item \textbf{\revtwo{JOB (SQLStorm).}}
\revtwo{We apply the same selection criterion for JOB: We select the 50 slowest queries by runtime under \texttt{olapbench}. As with the TPC-DS (SQLStorm) set, these queries are classified as medium to high complexity.}

\end{enumerate}

\head{Baselines} We compare \tool against two LLM-based approaches and three state-of-the-art rule-based query rewriting methods, two of which, \textsc{LLM-R\textsuperscript{2}} and \rbot, incorporate LLMs:
\begin{itemize}[leftmargin=0.5cm]
    \item \textbf{\baseline} employs a basic prompting technique illustrated in Figure~\ref{fig:basic_prompt} to generate candidate queries without any correction mechanism. 
        This baseline represents the straightforward application of out-of-the-box LLM.

        \item \revone{\textbf{LITHE}~\cite{dharwada2025query} is an LLM-based method that constructs a database-sensitive prompt by selecting the most appropriate  rule for the input query from a fixed set of six handcrafted rules. These rules target redundancy elimination and selectivity improvement.}

    \item \jie{\textbf{LLM-R\textsuperscript{2}} (VLDB 2025~\cite{li2024llm})  is an LLM-enhanced query rewrite system that can automatically select effective rules from a given set of rewrite rules to rewrite an input SQL query by leveraging a high-quality demonstration pool.}

    \item \textbf{R-Bot} (VLDB 2025~\cite{sun2024r}) is a query rewrite system that retrieves the most relevant rewrite evidences and employs an LLM-driven, step-by-step method to iteratively select and arrange rewrite rules based on retrieved evidence.

      \item \textbf{LearnedRewrite} (or \textbf{LR}, VLDB 2022~\cite{zhou2021learned}) is a state-of-the-art rule-based query rewriter~\cite{zhou2021learned} which uses Monte Carlo Tree Search to guide the rule-based rewriting process. LR uses rewrite rules from Calcite~\cite{begoli2018apache}, which is the most popular library for query rewriting rules.

\head{\revone{Comparison with QUITE~\cite{song2025quite}}} \revone{QUITE\cite{song2025quite} is also an LLM-based method for rewriting SQL queries beyond rules. However, the authors have not released code or several critical components (including the curated knowledge base and the LLM-based SQL corrector). Without access to these pieces, we cannot responsibly and confidently perform a direct experimental comparison. We therefore focus on high-level differences. QUITE models the query rewrite process as a Markov Decision Process and at each timestep it chooses a refinement action (e.g., join reordering, predicate pushdown) by matching the query against a curated knowledge base constructed from database documentation and Stack Overflow\cite{stackoverflow}. This constrains the search space to optimizations that are explicitly documented.
By contrast, \tool is not limited to a predefined action library. It can  apply optimization strategies that are implicitly encoded in the LLM  and accumulate novel patterns as NLR2s over time .}

% Since the authors have not published the code and some key components, e.g., curated knowledge base and the prompts for LLM-based SQL corrector, we cannot responsibly and confidently compare with their method. So we compare the high-level differences instead. It models the query rewrite process as a Markov Decision Process and at each timestep the agent chooses a refinement action (e.g., join reordering, predicate pushdown) based on the match with curated knowledge base, which is constructed from official database documentation and Stack Overflow~\cite{stackoverflow}. The issue is that this limits LLM's expressiveness and capabilities to apply patterns not explicitly listed in above resources to the query to achieve great speedups.

\end{itemize}

Note that while both \textsc{LLM-R\textsuperscript{2}} and \rbot utilize LLMs to enhance rule selection and sequencing, they do not explore beyond the space defined by the existing rule set. Therefore, we still classify them as rule-based approaches. We excluded ~\cite{wang2022wetune} from our experiments as it was not able to produce valid rewrites for any TPC-DS queries.

\head{Setup} \jie{Rule-based query rewriting methods produce the same rewrite for an input query across multiple runs. In contrast, due to the statistical nature of LLMs,} additional LLM runs increase the likelihood of uncovering new rewrite possibilities and thereby improving performance.  To ensure a fair comparison across approaches, we fix the number of iterations for each LLM-based method to four, generating four rewrite candidates per query. \baseline  uses the basic prompt (see Figure~\ref{fig:basic_prompt}) in all four iterations, without any correction or contextual enhancement.  \tool also uses the basic prompt in the first iteration, \revone{as the NLR2 Repository is initially empty and we do not use pre-seeded or external rules}. In subsequent iterations, it selectively incorporates relevant and beneficial NLR2s retrieved from the repository, if such rules are available, to guide the rewriting process more effectively. \revone{The semantic- and syntax-correction loops are each capped at 3 iterations. If, after these caps, a candidate still fails semantic correction or remains unparsable by the DBMS, we discard it and retain the original input $q$.} A method is considered successful on the given query $q$ if any of these four candidates is both equivalent to the original query and faster. If any method, rule-based or LLM-based, fails to produce an equivalent rewrite for the given query, we assign a speedup ratio of 1, indicating that the original query is used as a fallback.

% we generate four rewrites for each LLM-based method: both Baseline LLM and \tool share the first two candidate rewrites generated using the basic prompt (see Figure~\ref{fig:basic_prompt}), while the remaining two candidates differ—\tool incorporates the NLR2 with the highest estimated benefit as the hint, whereas Baseline LLM continues using the basic prompt. 

% A method is considered successful if any of these four candidates is both equivalent to the original query and faster. Additionally, if any method, rule-based or LLM-based, fails to produce an equivalent rewrite for a particular query, we deem the speedup ratio for that query as 1, indicating a fallback to the original query.

    % \item
    % \cut{ \textbf{Fusion} Computation Reuse via Fusion~\cite{bruno2022computation} is a technique for rewriting complex SQL queries in Amazon Athena. It focuses on minimizing computational redundancy by identifying and merging repeated operations, including instances where the common expressions are not exactly the same.}

% Both techniques had been evaluated on TPC-H or Leetcode~\cite{??} queries, which are much simpler than TPC-DS.

\begin{table}[tbp]
% \small
\footnotesize
\centering
\setlength{\tabcolsep}{6pt}
\def\arraystretch{1.1}
\vspace{-5pt}

\begin{subtable}[t]{0.48\textwidth}
\centering
\begin{tabular}{  c | r r r | r r r }
\toprule
& \multicolumn{3}{c|}{TPC-DS} 
& \multicolumn{3}{c}{JOB} \\
speedup & >2x & >10x & >50x & >1.2x & >2x & >10x \\
\midrule
\tool     & \makecell{25\\(25.3\%)} & \makecell{9\\(9.1\%)}  & \makecell{6\\(6.1\%)}  & \makecell{20\\(17.7\%)} & \makecell{8\\(7.1\%)}  & \makecell{3\\(2.7\%)} \\
\baseline & \makecell{18\\(18.2\%)} & \makecell{6\\(6.1\%)}  & \makecell{5\\(5.1\%)}  & \makecell{19\\(16.9\%)} & \makecell{7\\(6.2\%)}  & \makecell{3\\(2.7\%)} \\
\revone{\lithe} & \makecell{16\\(16.2\%)}    & \makecell{6\\(6.1\%)}  & \makecell{4\\(4.0\%)}    & \makecell{17\\(15.0\%)}    & \makecell{6\\(5.3\%)}    & \makecell{1\\(0.9\%)} \\
\textsc{LLM-R\textsuperscript{2}} 
          & \makecell{10\\(10.1\%)} & \makecell{6\\(6.1\%)}  & \makecell{5\\(5.1\%)}  & \makecell{9\\(8.0\%)}   & \makecell{2\\(1.8\%)}  & \makecell{1\\(0.9\%)} \\
\rbot     & \makecell{9\\(9.1\%)}   & \makecell{5\\(5.1\%)}  & \makecell{3\\(3.0\%)}  & \makecell{7\\(6.2\%)}   & \makecell{3\\(2.7\%)}  & \makecell{2\\(1.8\%)} \\
\lr       & \makecell{6\\(6.1\%)}   & \makecell{4\\(4.0\%)}  & \makecell{0\\(0\%)}    & \makecell{6\\(5.3\%)}   & \makecell{2\\(1.8\%)}  & \makecell{1\\(0.9\%)} \\
\bottomrule
\end{tabular}
\captionsetup{font=small}
\caption{Original workloads.}
% \inv
\label{tab:speedup_comparison}
\end{subtable}

\hfill

\begin{subtable}[t]{0.48\textwidth}
\centering
\begin{tabular}{  c | r r r | r r r }
\toprule
& \multicolumn{3}{c|}{TPC-DS (SQLStorm)} 
& \multicolumn{3}{c}{JOB (SQLStorm)} \\
speedup & >2x & >10x & >50x & >1.2x & >2x & >10x \\
\midrule
\tool     & \makecell{25\\(25.0\%)} & \makecell{13\\(13.0\%)}  & \makecell{3\\(3.0\%)}  & \makecell{7\\(14.0\%)} & \makecell{6\\(12.0\%)}  & \makecell{1\\(2.0\%)} \\
\baseline & \makecell{12\\(12.0\%)} & \makecell{9\\(9.0\%)}  & \makecell{3\\(3.0\%)}  & \makecell{4\\(8.0\%)} & \makecell{2\\(4.0\%)}  & \makecell{1\\(2.0\%)} \\
\revone{\lithe}    & \makecell{19\\(19\%)}    & \makecell{11\\(11.0\%)}  & \makecell{2\\(2.0\%)}    & \makecell{6\\(12.0\%)}    & \makecell{4\\(8.0\%)}    & \makecell{1\\(2.0\%)} \\
\textsc{LLM-R\textsuperscript{2}} 
          & \makecell{2\\(2.0\%)} & \makecell{1\\(1.0\%)}  & \makecell{0\\(0.0\%)}  & \makecell{2\\(4.0\%)}   & \makecell{2\\(4.0\%)}  & \makecell{0\\(0.0\%)} \\
\rbot     & \makecell{2\\(2.0\%)}   & \makecell{0\\(0.0\%)}  & \makecell{0\\(0.0\%)}  & \makecell{2\\(4.0\%)}   & \makecell{2\\(4.0\%)}  & \makecell{0\\(0.0\%)} \\
\lr       & \makecell{4\\(4.0\%)}   & \makecell{1\\(1.0\%)}  & \makecell{0\\(0\%)}    & \makecell{3\\(6.0\%)}   & \makecell{1\\(2.0\%)}  & \makecell{0\\(0.0\%)} \\
\bottomrule
\end{tabular}
\captionsetup{font=small}
\caption{\revtwo{SQLStorm-generated workloads.}}
\label{tab:speedup_comparison_sqlstorm}
\end{subtable}

\captionsetup{font=small,name=Table}
\caption{Speedup comparison beyond thresholds between all methods. Upper: original TPC-DS and JOB. Lower: SQLStorm-generated TPC-DS and JOB. We report absolute counts and percentages to reflect the different workload sizes.}
\label{tab:speedup_all}
\vspace{-15pt}
\end{table}

\subsection{Comparison with \revone{LLM-based Methods}}
\label{sec:expr:llmbaseline}

Table~\ref{tab:speedup_all} compares the fraction of queries that each method speeds up beyond several thresholds. Before discussing individual results, we note an important trend: queries in JOB are generally less complex than those in TPC-DS, both in the original benchmarks and in the SQLStorm-generated variants. In original workload, JOB queries average 278.81 tokens, whereas TPC-DS queries average 441.34 tokens (a 58\% increase). Although JOB queries can involve many joins and are therefore challenging for join ordering, their overall structure is simpler: they contain fewer nested subqueries, window functions, and set operations than TPC-DS. \revtwo{This gap persists in the SQLStorm workloads. Even when SQLStorm generates new queries with medium to high complexities over both schemas, the JOB-derived queries remain constrained by narrower, more normalized tables and by the absence of complex cross-channel or temporal comparisons. This difference in query complexity translates to different optimization potential. In TPC-DS and its SQLStorm variants, \tool can optimize roughly 25\% of queries by at least 2x, whereas in JOB and its SQLStorm variant, only about 7\%--12\% of queries reach that level of improvement.}

Overall, \tool delivers the strongest improvements across all benchmarks, and this advantage is most pronounced on the more complex TPC-DS workloads. LLM-based methods outperform rule-based query rewriting approaches by notable margins. On the original TPC-DS benchmark, \tool speeds up 25 queries by at least $2$x and 9 queries by at least $10$x. By comparison, \baseline speeds up 18 and 6 queries at those thresholds, and \lithe speeds up 16 and 6 queries. \revtwo{On the SQLStorm-generated TPC-DS workload, \tool again leads: it speeds up 25 queries by at least $2$x and 13 queries by at least $10$x, compared to 12 and 9 for \baseline and 19 and 11 for \lithe. On JOB and JOB (SQLStorm), all three methods show more limited gains, since these queries offer less optimization opportunity. \tool still exhibits a slight edge, and the gap between methods is less pronounced than on TPC-DS.}

\revtwo{An interesting finding is that \baseline outperforms \lithe on the standard TPC-DS benchmark, but the ranking flips on TPC-DS (SQLStorm); we observe the same trend on JOB vs.\ JOB (SQLStorm). This suggests that \baseline benefits from the LLM’s prior exposure to well-known public benchmarks and can often reproduce established optimizations there. On newly generated, previously unseen queries (e.g., TPC-DS (SQLStorm)), however, \baseline both struggles to suggest a good rewriting direction and more frequently produces invalid SQL. \lithe is more robust in that setting because it explicitly chooses one of six handcrafted optimization rules (e.g., redundancy removal or increased selectivity) and conditions the LLM on that rule plus an example rewrite; this often gives the model a strong starting direction. But \lithe has two structural limitations: it cannot repair incorrect rewrites, and its handcrafted rule set cannot evolve as new workloads arrive. \tool addresses both issues: it uses counterexample-guided iterative correction to fix invalid or semantically inequivalent rewrites, and it transfers knowledge via NLR2s that accumulate and improve over time, increasing the likelihood that the generated rewrite is both correct and faster.}

Table~\ref{tab:equivalence} compares performance of \tool and \baseline in terms of the correctness of generated rewrites. Both methods are nearly perfect on JOB, where query complexity is low. On TPC-DS benchmark, \tool produces \revtwo{at least one} equivalent rewrites for 85.9\% of queries, outperforming \baseline’s 74.7\%. Moreover, 70\% of \tool’s generated rewrites are correct, compared to only 51.8\% for \baseline. Notably, approximately one-third of \baseline’s rewrites fail to execute due to minor syntax errors or fabricated table/column names, while only 13.3\% of \tool’s rewrites suffer from such issues, highlighting the effectiveness of our iterative correction module. These correctness differences matter for performance. Because \baseline only generates a correct rewrite about half the time, its ability to suggest a promising optimization direction does not always translate into a usable speedup. Table~\ref{tab:speedup_all} shows that \tool reliably finds more faster rewrites than \baseline within the same four rewrite iterations. This does not imply that \baseline is inherently incapable of finding faster rewrites for them; rather, it may require substantially more iterations and LLM API calls to do so. However, the number of additional runs needed is unpredictable, making the approach less practical in real-world settings. In contrast, \tool uses NLR2s to carry forward rewrite knowledge from previously optimized queries and incorporates performance feedback into subsequent prompts, increasing the likelihood of discovering effective rewrites in fewer iterations.

While \tool and \baseline achieve similar speedup for more than half of these representative queries, there are a few interesting cases where \tool significantly outperforms \baseline. These cases highlight the value of incorporating NLR2s in \tool.

\begin{table}[!t]
\footnotesize

\vspace{-3pt}
\centering
\setlength{\tabcolsep}{3pt}
\def\arraystretch{1.2}
\begin{tabular}{ p{23pt} | p{120pt} | r  r   }
\toprule
 &
& \tool
& \baseline
\\ \midrule
\multirow{5}{*}{TPC-DS} 
&  {\% of input queries with \revtwo{$\geq1$ equiv. rewrites}} 
& {85.9\%}
& {74.7\%} 
\\ 
& {\% of rewrites that are equivalent} 
& {70.0\%}
& {51.8\%}
\\ 
& {\% of rewrites that are inequivalent}
& {13.4\%}
& {13.1\%}  

\\ 

& {\% of rewrites that are undetermined}
& {3.3\%}
& {2.0\%}  
\\ 
& {\% of rewrites that cannot be executed}
& {13.6\%} 
& {33.1\%} 

\\ \midrule \midrule 

\multirow{5}{*}{JOB}
&  {\% of input queries with \revtwo{$\geq1$ equiv. rewrites}} 
& {100\%}
& {100\%} 
\\ 
& {\% of rewrites that are equivalent} 
& {99.8\%}
& {98\%}
\\ 
& {\% of rewrites that are inequivalent}
& {0\%}
& {0.7\%}  

\\ 

& {\% of rewrites that are undetermined}
& {0\%}
& {0\%}  
\\ 
& {\% of rewrites that cannot be executed}
& {0.2\%} 
& {1.5\%} 
\\ \bottomrule
\end{tabular}
\captionsetup{font=small,name=Table}
\caption{Equivalence checking results on original TPC-DS and JOB benchmark for \tool and \baseline. \revtwo{For each input query, both methods generate four candidate rewrites. The first row shows the percentage of input queries for which at least one equivalent rewrite was found. The remaining rows classify all generated rewrites (across all four attempts) by outcome: equivalent, inequivalent, undetermined, or failed to execute.}}
\vspace{-12pt}
\label{tab:equivalence}
\end{table}

\head{Case 1: Both can optimize Q4, Q11, Q74, but \tool achieves much better speedup}

\noindent\tool infers from the query plan that Q4, Q11, and Q74 likely share similar bottlenecks, specifically, issues such as ``\textit{Excessive nested loop joins on CTE scans}'' and ``\textit{Disk‐based sort and aggregate operations on large fact tables}''. Consequently, these queries can benefit from analogous optimization strategies. More specifically, the NLR2 ``\textit{Replace multiple UNION subqueries with separate aggregation subqueries using CASE expressions}'' is derived from Q74, while 
the NLR2 ``\textit{Split aggregated computations by filtering on the specific period in separate subqueries}'' originates from Q11. Both of these transformations yield a higher speedup for Q4 compared to its original effective NLR2: ``\textit{Replace multiple self-joins with pivoted aggregations}'' (18x, 7x vs. 3.5x). Notably, the NLR2 from Q11 produces more consistent improvements for Q4, as the transformation is relatively straightforward and encounters fewer issues of inequivalence.

\head{Case 2: \baseline cannot find equivalent rewrites for Q23 while \tool can}

\noindent LLMs are trained on vast datasets and accumulate substantial domain knowledge, which often enables them to perform intelligent optimizations. However, in some cases an LLM’s chosen strategy can be flawed, consistently yielding inequivalent outputs. For example, Listing~\ref{listing:inequivalent_nlr2} shows two queries (irrelevant parts omitted for clarity): on the left, TPC-DS Q23, and on the right, a candidate rewrite generated by \baseline, which consistently produces a similar query pattern over multiple rounds. Although the rewrite appears plausible, a subtle difference leads to divergent result sets. Specifically, the rewritten query joins \texttt{web\_sales} with a common table expression (CTE) such that qualifying items may match more than once, duplicating rows and inflating the final sum. In contrast, the corresponding \tool rewrite employs an IN predicate on \texttt{item\_sk} from \texttt{frequent\_items}. Even if the CTE contains duplicate rows, the \sqlin clause merely checks for membership and does not multiply matches. This difference in join behavior explains why the two queries yield markedly different totals when executed at scale. \tool adopts the NLR2 ``\textit{Pre-aggregate the  CTE  so it is computed once and returns fewer rows}'' from Q95, achieving 3.34x speedup. This case highlights the importance of exercising explicit control over rewriting directions through the application of NLR2s, rather than relying on LLMs as black boxes. It also emphasizes the role of rule separation in facilitating effective and accurate knowledge transfer across queries.

\begin{listing}[h]
  \tiny  % Reduce font size (use \footnotesize, \scriptsize, or even \tiny)
  \centering
  % \inv
  \begin{minipage}[t]{0.49\linewidth}  % First half
    \begin{tcolorbox}[colback=gray!5, colframe=gray!40, boxsep=2mm, left=6pt, right=5pt, top=0.5mm, bottom=0.5mm]
      \begin{minted}[
          frame=none,
          framesep=2mm,
          linenos,
          numbersep=2mm,
          escapeinside=||,
          baselinestretch=1.0, 
          fontsize=\tiny
      ]{sql}
            ...
select sum(sales)
...
from ...
|\textcolor{blue}{$web\_sales$}| 
where ws_item_sk in
   (select item_sk
        FROM |\textcolor{blue}{$frequent\_items$}|)
   ...
      \end{minted}
    \end{tcolorbox}
    \vspace{-1mm}
    \caption*{(a) TPC-DS Q11 Original}  % Sub-caption inside the same listing
  \end{minipage}
  \hfill
  \begin{minipage}[t]{0.49\linewidth}  % Second half
    \begin{tcolorbox}[colback=gray!5, colframe=gray!40, boxsep=2mm, left=6pt, right=5pt, top=0.5mm, bottom=0.5mm]
      \begin{minted}[
          frame=none,
          framesep=2mm,
          linenos,
          numbersep=2mm,
          escapeinside=||,
          baselinestretch=1.0, 
          fontsize=\tiny
      ]{sql}

      ...
select sum(sales)
...
from ...
|\textcolor{blue}{$web\_sales$}| as ws
   join |\textcolor{blue}{$frequent\_items$}| as f 
   on ws.ws_item_sk = f.item_sk
   ...
      \end{minted}
    \end{tcolorbox}
    \vspace{-1mm}
    \caption*{(b) TPC-DS Q11 Rewritten}
  \end{minipage}
  \vspace{-1mm}
  \captionsetup{font=small,name=Listing}
  \caption{Left: Original TPC-DS Q23 query. Right: A rewrite of Q23 generated by \baseline that is inequivalent.}
  \inv
  
  \label{listing:inequivalent_nlr2}
\end{listing}

\subsection{Comparison with Rule-based Methods}
\label{sec:expr:traditionalbaseline}

Analysis of Table.~\ref{tab:speedup_all} shows that \tool consistently surpasses the
three rule-based baselines in terms of the number of queries optimized, regardless of the desired speedup threshold, on all benchmarks. Other LLM-based methods also consistently outperform rule-based methods with smaller margins than \tool.
We first consider the standard public benchmarks (Table~\ref{tab:speedup_comparison}). On the original TPC-DS workload, \tool identifies substantially more optimization opportunities than the rule-based baselines: it achieves at least 2x speedup on 25 queries, compared to 10, 9, and 6 queries for \textsc{LLM-R\textsuperscript{2}}, \rbot, and \lr, respectively. The advantage persists at higher thresholds (e.g., 10x speedup), and we observe a similar pattern on JOB. These results suggest that \tool already outperforms established rule-based methods on well-studied public benchmarks.

\noindent\revtwo{SQLStorm workloads (Table~\ref{tab:speedup_comparison_sqlstorm}) yield two key observations:}

\head{\revtwo{Observation 1}}
\revtwo{The performance gap between \tool and the rule-based baselines widens on TPC-DS (SQLStorm).
    All three rule-based baselines (\textsc{LLM-R\textsuperscript{2}}, \rbot, and \lr) are built on top of Calcite and ultimately rely on Calcite’s rewrite rules. Their differences lie in rule selection and ordering, but they inherit the same two Calcite limitations: (1) \emph{Parser coverage:} The Calcite parser and validator do not fully support several constructs that appear in SQLStorm-generated  TPC-DS queries, such as recursive CTEs, \texttt{LENGTH}, etc. As a result, these systems only emit rewrites for roughly half of the input queries. (2) \emph{Rule coverage:} Even when Calcite can parse the query, its fixed rule set often fails to match the structure of these queries. SQLStorm's queries are not hand-polished benchmark SQL; they contain noisy nesting, ad hoc join predicates, and unseen operator combination that does not line up cleanly with Calcite’s stock transformations (e.g., predicate pushdown). In these cases, the systems may not show meaningful improvement. In contrast, \tool does not depend on Calcite’s rule inventory. It can generate and iteratively refine rewrites for these same SQLStorm queries, which explains the widening gap on TPC-DS (SQLStorm).}

\head{\revtwo{Observation 2}}
\revtwo{On SQLStorm workloads, the ``LLM-enhanced'' rule-based systems degrade to the point that they can be outperformed by \lr, which does not use an LLM at all. \textsc{LLM-R\textsuperscript{2}} selects a few high-quality demonstration rewrites from a curated library using a learned retriever, and \rbot retrieves offline ``rewrite recipes'' mined from past queries. These strategies work on TPC-DS and JOB because the benchmark queries follow canonical patterns that are well represented in those demonstration pools. Under SQLStorm, however, the generated queries deviate substantially from those canonical patterns, so the retrieved demonstrations and recipes are often no longer relevant and provide poor guidance. \tool, by contrast, is not limited to a static set of expert demonstrations or a fixed rule inventory. It employs iterative correction and accumulates transferable rewrite knowledge from prior queries, enabling it to adapt to new query forms without manual rule engineering.}

\revtwo{\tool outperforms existing baselines both on the original TPC-DS / JOB benchmarks and on the SQLStorm-generated workloads. Its advantage becomes larger precisely in the regimes where fixed-rule systems provide poor coverage. If new rules were distilled from \tool's successful rewrites on complex queries and then incorporated into Calcite, the rule-based methods could  be strengthened and extended to a broader class of queries.}

\head{\revtwo{A deeper look at rules discovered by \tool}}
\revtwo{Besides standard Calcite rules (e.g., predicate pushdown, projection pruning), several papers propose more aggressive rewrite patterns for analytical workloads. Generalized Sub-Query Fusion~\cite{sarthi2020generalized} eliminates redundant I/O by fusing multiple similar subqueries into a single shared scan/aggregation. Super-operators~\cite{leeka2019incorporating} collapse multi-branch aggregates and unions into a single streaming physical operator. The Spark engine in Azure Synapse~\cite{modi2021new} rewrites plans to reuse shuffles, push down filters and partial aggregates, and inject runtime partition pruning to reduce data movement. Athena~\cite{bruno2022computation} detects repeated expensive subplans in a query and fuses them so they are computed once and shared by all consumers. In Table~\ref{tab:pattern-summary}, for each major rewrite pattern discovered by \tool we ask: (1) does it align with any of these known patterns, at the idea level; and (2) is there an exact structural match already reported in prior work?}

% [leftmargin=0.5cm]

\noindent\revone{Representative optimization patterns discovered by \tool:}\footnote{In Appendix~A, we present query pairs for each pattern and detail how they differ from prior work.}
\begin{enumerate}[leftmargin=0.5cm]
\item  \revone{Pivot with conditional aggregation instead of UNION-then-self-join.}
\item \revone{Replace multiple \texttt{EXISTS} conditions with a precomputed set via joins.}
\item \revone{Rewrite \texttt{COUNT(*)>0} as \texttt{EXISTS} (aggregate $\rightarrow$ semi-join; enables short-circuiting).}
\item \revone{Replace a bounded self-recursive CTE with a non-recursive base query \texttt{CROSS JOIN}ed to a small numbers table.}
\item \revone{Lift a per-key summary (e.g., item-level totals) into one CTE and drive all branches from it.}
\item \revone{Consolidate repetitive subqueries into a single aggregation subquery with conditional expressions.}
\item \revone{Decorrelate \texttt{IN} by joining a \texttt{SELECT DISTINCT} key,... mapping table instead of re-checking the subquery. }
\end{enumerate}

\begin{table}[tbp]
\inv
\small
\centering
\setlength{\tabcolsep}{6pt}
\def\arraystretch{1.2}

\begin{tabular}{ l | c c c }
\toprule
Pattern& Covered queries & Similar in idea & Exact match \\
\midrule
1 & 4, 11, 74 & ~\cite{leeka2019incorporating}, ~\cite{bruno2022computation} & No \\
2 & 10, 35, 69 & ~\cite{begoli2018apache} & No \\
3 & 41 & ~\cite{begoli2018apache} & No \\
4 & 23998* &  & No \\
5 & 22722* & ~\cite{sarthi2020generalized}, ~\cite{bruno2022computation} & No \\
6 & 61, 88 & ~\cite{sarthi2020generalized}, ~\cite{bruno2022computation} & Yes,~\cite{bruno2022computation}  \\
7 & 28982*, 34389* & ~\cite{begoli2018apache} & Yes,~\cite{begoli2018apache}  \\
\bottomrule
\end{tabular}
\captionsetup{font=small,name=Table}
\caption{\revtwo{Summary of rewrite patterns discovered by \tool. ``Covered queries'' lists the queries that exhibit each pattern (queries with * are from TPC-DS SQLStorm; others are from standard TPC-DS). ``Similar in idea'' notes prior work that proposes a conceptually related optimization. ``Exact match''indicates that the same structural rewrite has been reported before.}  }
\inv
\label{tab:pattern-summary}
\end{table}

% \begin{enumerate}[leftmargin=0.5cm]

% \item  Pivot with conditional aggregation instead of UNION-then-self-join.
% \item Rewrite \texttt{COUNT(*)>0} as \texttt{EXISTS} (aggregate $\rightarrow$ semi-join; enables short-circuiting).
% \item Consolidate repetitive subqueries into a single aggregation subquery with conditional expressions.

% \end{enumerate}

\subsection{Detailed Analysis and Ablation Study}
\label{sec:expr:ablation}

In this section, we conduct a detailed ablation study 
    to validate our design decisions in  \tool and the contribution of each component therein towards the overall performance. We focus primarily on the \revtwo{original} TPC-DS benchmark, as its complexity provides a challenging test of our system's capabilities. The results, summarized in Table~\ref{tab:ablation}, highlight the impact of individual components on LLM runtime overhead, monetary cost, and the percentage of queries achieving various levels of speedup.
    % To control for variables, we still adopt a setup in which our NLR2 repository is initialized with two LLM runs using the basic prompt and \tool incorporates the NLR2 as a hint for the remaining two candidate rewrites,.

\begin{table*}[!t]
\small
\centering
\setlength{\tabcolsep}{10pt}
\def\arraystretch{1.1}
\vspace{-5pt}
\begin{tabular}{ | c | r r r r r |  }
\toprule
& \multicolumn{5}{c|}{TPC-DS} \\
 & >2x & >10x & >50x  & avg. time(s) & avg. cost(\$)  \\
\midrule
\tool                           & \cntpct{25}{25.3} & \cntpct{9}{9.1}   & \cntpct{6}{6.1}   & 151.82 & 0.129 \\
\tool without grouping NLR2s    & \cntpct{25}{25.3} & \cntpct{9}{9.1}   & \cntpct{6}{6.1}   & 385.47 & 0.310 \\
\tool without bottleneck analysis & \cntpct{19}{19.2} & \cntpct{6}{6.1}   & \cntpct{5}{5.1}   & 150.30 & 0.124 \\
\tool with gpt-4o               & \cntpct{14}{14.1} & \cntpct{5}{5.1}   & \cntpct{5}{5.1}   & 105.40 & 0.117 \\
\tool with gpt-4-mini           & \cntpct{2}{2.0}   & \cntpct{1}{1.0}   & \cntpct{0}{0.0}   & 301.20 & 0.016 \\
\bottomrule
\end{tabular}
\captionsetup{font=small,name=Table}
\caption{Ablation study: impact of \tool's design decisions on LLM runtime, cost, and the share of queries achieving various speedups.}
\label{tab:ablation}
\vspace{-15pt}
\end{table*}

\head{Plan-based dominant NLR2 identification} This component plays a critical role in reducing query execution overhead, although its impact is not directly reflected in Table~\ref{tab:ablation}. Without plan-based dominant NLR2 identification, we would need to execute each rewrite—generated by applying a single NLR2—to measure its runtime, which can be prohibitively expensive for long-running queries. 

\head{NLR2 grouping via LLM}
We collect an average of 5.08 NLR2s per input query after applying grouping, compared to 13.24 NLR2s per query without grouping. Although we only re-evaluate the impact of an NLR2 when the associated performance improvement exceeds a predefined threshold, grouping still significantly reduces both monetary and time costs. Without grouping, the average LLM runtime increases by 153.9\%, and the monetary cost rises by 140.3\%,  demonstrating that NLR2 grouping's impact on costs.

\head{Query performance bottleneck analysis} 
We compare two strategies: (1) apply the top NLR2s globally, ignoring query similarity; and (2) our default, which retrieves candidates by bottleneck similarity. The global strategy performs essentially the same as \baseline, indicating that ``best overall'' strategy rarely transfer. Bottleneck-aware retrieval, in contrast, delivers clear gains: effective reuse hinges on aligning NLR2s with performance bottlenecks.

% We evaluate two strategies for identifying queries with similar performance bottlenecks: (1) applying the top beneficial NLR2s regardless of query similarity, and (2) our default approach, which incorporates bottleneck-aware retrieval. Experimental results show that the first strategy performs nearly identically to \baseline, suggesting that indiscriminately applying high-performing NLR2s provides little benefit. This underscores the importance of relevance—effective reuse of NLR2s depends on accurately identifying queries with comparable performance bottlenecks.

\head{Choice of the LLM}
\tool's performance critically depends on LLM's reasoning capabilities. We evaluate three models: OpenAI’s \texttt{o3-mini} (reasoning-oriented), \texttt{gpt-4o} (balanced), and \texttt{gpt-4o-mini}(most cost-efficient).  \tool powered by \texttt{gpt-4o} is noticeably outperformed by the \texttt{o3-mini} variant. We decompose \tool’s workflow into two stages: (1) discovering effective rewrites and summarizing NLR2s, and (2) applying high-quality NLR2s to guide rewrites. We observe that \texttt{gpt-4o} struggles primarily in stage 1. However, when \tool with \texttt{gpt-4o} is supplied with NLR2s previously discovered by \texttt{o3-mini}, the number of queries achieving a speedup $\geq$ 10× increases from 5 to 8,  further demonstrating the effectiveness of NLR2-guided rewriting.
% \texttt{gpt-4o-mini} rarely produces useful rewrites.

% \tool’s gains track the model’s reasoning strength. \texttt{o3-mini} (reasoning-oriented) consistently finds stronger rewrites and better NLR2s, outperforming \texttt{gpt-4o}; \texttt{gpt-4o-mini} rarely produces useful rewrites. Decomposing \tool into (i) discovery/summarization of NLR2s and (ii) application shows where the gap arises: \texttt{gpt-4o} underperforms in (i) but applies good NLR2s well—when seeded with \texttt{o3-mini}’s NLR2s, its count of >10× speedups rises from 5 to 8. Practical takeaway: use the strongest model for discovery, then a cheaper model can reliably apply the learned NLR2s.

\head{\revone{Parameter sensitivity analysis}}

\begin{table}[tbp]
\small
\centering
\begin{tabular}{l*{4}{cc}}
\toprule
& \multicolumn{2}{c}{$k{=}1$}
& \multicolumn{2}{c}{$k{=}2$}
& \multicolumn{2}{c}{$k{=}3$}
& \multicolumn{2}{c}{$k{=}4$} \\
\cmidrule(lr){2-3}
\cmidrule(lr){4-5}
\cmidrule(lr){6-7}
\cmidrule(lr){8-9}
& eq & fast
& eq & fast
& eq & fast
& eq & fast \\
\midrule
No correction
& 52 & 5
& 67 & 8
& 70 & 12
& 74 & 21 \\
Semantic only
& 54 & 5
& 68 & 10
& 71 & 14
& 75 & 21 \\
\textbf{Semantic + syntax}
& \textbf{70} & \textbf{5}
& \textbf{79} & \textbf{10}
& \textbf{85} & \textbf{16}
& \textbf{86} & \textbf{28} \\
\bottomrule
\end{tabular}
\captionsetup{font=small,name=Table}
\caption{\revone{Cumulative count of queries with (i) at least one equivalent rewrite (\textit{equiv}) and (ii) at least one equivalent rewrite that is at least 20\% faster than the original (\textit{faster}), as the candidate budget increases ($k{=}1\ldots4$), under different correction strategies.}}
\label{tab:correction_ablation_new}
\inv
\end{table}

\noindent\revone{
\emph{\textbf{Budget:}} maximum candidates per query \(K=4\); correction caps \(I_{\mathrm{sem}}=I_{\mathrm{syn}}=3\) (semantic and syntax loops). For the TPC-DS workload, the average iterations is 0.30 (semantic) and 0.76 (syntax). The per-query correction iteration distributions are:}
\begin{itemize}
    \item \revone{Semantic: 0: 76.72\%, 1: 19.02\%, 2: 1.64\%, 3: 2.62\%.}
    \item \revone{Syntax: 0: 60.66\%, 1: 18.36\%, 2: 5.25\%, 3: 15.74\%.}
\end{itemize}
\revone{We vary the candidate budget $k \in\{1,2,3,4\}$. Table~\ref{tab:correction_ablation_new} reports the \emph{cumulative} number of queries that achieve (i) $\geq 1$ equivalent rewrite and (ii) $\geq 1$ equivalent rewrite that is at least 20\% faster, as $k$ increases (higher \(k\) means more candidates attempted per input query), comparing: (i) no correction, (ii) semantic-only correction, and (iii) semantic\,+\,syntax correction.} \revone{Table~\ref{tab:correction_ablation_new} shows that correction loops markedly reduce the effective error rate,  with \emph{semantic + syntax} outperforming the other settings at every \(k\). Larger \(k\) improves coverage with diminishing returns and proportionally higher cost.}

\noindent\revone{\emph{\textbf{Desired speedup threshold:}}
We set $\theta$ to 1.2x in our implementation. Increasing $\theta$ reduces coverage  but raises the average speedup among accepted rewrites. With a lower $\theta$, we keep more rewrites — including many that are only modestly better.}

\begin{table}[H]
\small
\centering
\begin{tabular}{lccc}
\toprule
Desired speedup threshold $\theta$ & 1.2x & 2.0x & 5.0x \\
\midrule
Coverage (\%)       & 28.3 & 23.2 & 13.1 \\
Geom.\ mean speedup & 11.1x & 17.0x & 72.2x \\
\bottomrule
\end{tabular}
\captionsetup{font=small,name=Table}
\caption{\revone{Coverage and geometric mean speedup for desirable speedup thresholds $\theta \in \{1.2x,1.5x,2.0x\}$. 
Coverage(\%) is the fraction of queries with $\ge\theta$ speedup; geom.\ mean is over covered queries.}}
\label{tab:theta_sweep}
\inv
\end{table}

\vspace{-5mm}

\head{\revthree{Counterexample-guided iterative correction}} 
\revthree{We compare our counterexample-based semantic check to Miniature \& Mull from LLM-SQL-Solver~\cite{zhao2023llm} on random TPC-DS rewrites (no correction). Our prompt yields 29\% equivalent rewrites before syntax correction with a 6\% false-positive rate; Miniature \& Mull achieves 26\% equivalents with 22\% false positives. Thus our prompt better aligns with the iterative correction task—more verified equivalents and far fewer false positives.}

\subsection{Time and Monetary Cost}
% \subsection{Cost analysis of LLM}
\label{sec:expr:llmcost}

While \tool’s time and API costs are a one-time, amortized expense for recurring workloads (see Target workloads in \S\ref{sec:intro}), we still quantify them to ensure they remain modest. Using OpenAI’s \texttt{o3-mini} to generate a rewrite for a TPC-DS query—including suggestion, semantic correction, and syntactic correction—costs on average \$0.029 and takes 36.15 s. For comparison, \texttt{gpt-4o} and \texttt{gpt-4o-mini} average (\$0.026, 25.1 s) and (\$0.004, 71.71 s), respectively. As a reasoning-focused model, \texttt{o3-mini} consumes more tokens, increasing cost but often yielding more logically sound, well-structured rewrites for complex SQL transformations. Auxiliary LLM steps (e.g., grouping NLR2s and selecting the bottleneck) add negligible overhead. Overall, total cost scales approximately linearly with the number of candidate rewrites generated by the pipeline.

\begin{table}[h]
% \small
\footnotesize
\centering
\setlength{\tabcolsep}{6pt}
\def\arraystretch{1.1}
\vspace{-5pt}
\begin{tabular}{  c | r r r | r r r }
\toprule
& \multicolumn{3}{c|}{time (s)} 
& \multicolumn{3}{c}{monetary cost (\$)} \\
stage & suggest & \makecell{semantic\\corr.} & \makecell{syntax\\corr.} & suggest & \makecell{semantic\\corr.} & \makecell{syntax\\corr.} \\
\midrule
o3-mini     & \makecell{5.8\\(16.0\%)} & \makecell{20.8\\(57.6\%)}  & \makecell{9.6\\(26.5\%)}  & \makecell{7.8e-03\\(26.4\%)} & \makecell{1.5e-02\\(52.2\%)}  & \makecell{6.3e-03\\(21.3\%)} \\
gpt-4o & \makecell{1.5\\(5.8\%)} & \makecell{22.1\\(87.9\%)}  & \makecell{1.6\\(6.3\%)}  & \makecell{6.4e-03\\(23.9\%)} & \makecell{1.9e-02\\(70.1\%)}  & \makecell{1.6e-03\\(6.0\%)} \\
gpt-4o-mini & \makecell{1.9\\(2.6\%)} & \makecell{66.8\\(93.1\%)}  & \makecell{3.1\\(4.3\%)}  & \makecell{4.0e-04\\(11.2\%)}  & \makecell{3.1e-03\\(84.6\%)}   & \makecell{1.5e-04\\(4.2\%)} \\
% \rbot       & \makecell{9\\(9.1\%)}  & \makecell{5\\(5.1\%)}   & \makecell{3\\(3.0\%)}   & \makecell{7\\(6.2\%)}  & \makecell{3\\(2.7\%)}   & \makecell{2\\(1.8\%)} \\
% \lr       & \makecell{6\\(6.1\%)}  & \makecell{4\\(4.0\%)}   & \makecell{0\\(0\%)}   & \makecell{6\\(5.3\%)}  & \makecell{2\\(1.8\%)}   & \makecell{1\\(0.9\%)} \\
\bottomrule
\end{tabular}
\captionsetup{font=small,name=Table}
\caption{\small LLM cost analysis: the monetary and time cost of generating one rewrite candidate for a TPC-DS query, including all three stages--- rewrite suggestions, semantic correction, and syntactic correction.}
\inv
\label{tab:llm_cost}
\vspace{-10pt}
\end{table}

 % The breakdown of the monetary and time cost is presented in Table~\ref{tab:llm_cost}. Across all three models, semantic correction phase accounts for the highest cost in terms of both monetary and time cost. However, when using \texttt{o3-mini}, a larger portion of time is also spent on the candidate suggestion and syntax correction phases. This is because \texttt{o3-mini}, with its stronger reasoning capabilities, tends to better understand the input query and generate more aggressive rewrites. While this increases the likelihood of discovering high-quality transformations, it also lowers the chance of producing equivalent rewrites on the first attempt, thereby requiring additional iterations in the syntax correction phase. In contrast, the two lower-capacity models exhibit a more conservative rewriting style, leading to fewer iterations but also fewer opportunities for substantial performance improvement.

 Table~\ref{tab:llm_cost} shows that semantic correction dominates both time and cost across models. With \texttt{o3-mini}, more time also goes to candidate suggestion and syntax correction: its stronger reasoning yields more aggressive rewrites (more chances for big wins) but requires extra iterations to ensure equivalence. By contrast, \texttt{gpt-4o} and \texttt{gpt-4o-mini} rewrite more conservatively, needing fewer iterations but offering fewer opportunities for substantial speedups.

\revone{For equivalence checking, we set a 60 sec timeout for the tester and 5 sec for the verifier. Performance–evaluation time varied by query from seconds to hours. We also terminated execution early when a candidate failed to exceed the original query by the desired speedup threshold $\theta$.}

\cut{
\lr~\cite{zhou2021learned} and \textsc{LLM-R\textsuperscript{2}}~\cite{li2024llm} both integrate
Apache Calcite~\cite{begoli2018apache} as the underlying rewrite engine and learn to select from Calcite’s rewrite rules to maximize their benefits. Nonetheless, their ability to rewrite SQL queries is inherently limited by the scope and expressiveness of Calcite's rule set.}

\section{Discussion}
\label{sec:discussion}

Determining the equivalence of rewrites is a common challenge for all LLM-based approaches. Even the most powerful SQL verifier to date, SQLSolver~\cite{ding2023proving}, can verify only 18.9\% of the generated TPC-DS rewrites. The verification rate is significantly higher for the JOB benchmark at 74.4\%, yet it still fails to verify all rewrites. Moreover, as queries grow more complex, it becomes increasingly difficult for testers to spot inequivalent queries. To ensure correctness, all rewrites reported in the evaluation section have been manually inspected.
Although verifying equivalence can be resource-intensive, the potential benefit of identifying a significantly more efficient rewrite often justifies the cost. 
% For example, several TPC-DS queries have runtimes exceeding one hour, whereas \tool is able to generate equivalent rewrites that reduce execution time by more than an order of magnitude. As a reference point, a Snowflake ``Small'' virtual warehouse costs about \$2.00 per hour. 
The modest investment in LLM usage can yield substantial savings by reducing data warehouse compute costs, especially for long-running queries.
In the context of recurring query execution, \tool can serve as an automated system to propose query pairs that uncover valuable traditional rewrite rules. Consequently, \tool is not designed to replace rule-based methods; rather, it leverages the capabilities of LLMs to identify promising rules that can be integrated into existing database. \revtwo{\tool returns verified  query pairs that deliver measured speedups together with the associated NLR2. A rule distilled from only one such pair is brittle and often fails to generalize—it overfits incidental syntax choices and misses necessary preconditions. We can aggregate multiple verified pairs tied to the same NLR2, and extract the common AST/plan transformation plus explicit guards/preconditions. As more pairs accumulate, the resulting pattern-matching rule becomes more general  while remaining safe.\footnote{In Appendix~B, we present a concrete example showing how NLR2 clusters query pairs with shared issues, allowing us to extract robust rewrite rules suitable for database implementation.}}

% Microsoft: The need for global optimization decisions is further motivated by the observation that business-critical jobs in analytics clusters are typically recurrent. Instances of the same job are issued periodically over new batches of data, e.g., hourly, daily, or weekly. In fact, over 60% of the jobs in our clusters are recurrent, with the majority of them being submitted daily.
\vspace{-1mm}

\section{Related Work}
\label{sec:related}

In this section, we summarize the most relevant related work.

\head{Query rewriting} Manual query rewriting does not scale and is prone to error. 
Much of the work in automated query rewriting is rule-based ~\cite{pirahesh1992extensible,graefe1987rule,ahmed2006cost,levy1997query,muralikrishna1992improved,seshadri1996cost}.  LearnedRewrite~\cite{zhou2021learned} builds on Calcite’s classic rewrite rules and explores how to find the optimal order of applying rules. \revtwo{More recent systems propose new rewrite rules tailored to analytical workloads~\cite{sarthi2020generalized,leeka2019incorporating,modi2021new,bruno2022computation}, as discussed in Section ~\ref{sec:expr:llmbaseline}.} The emergence of LLMs has inspired LLM-enhanced rule-based approaches. \rbot~\cite{sun2024r} combines LLMs with evidence retrieval and self-reflection mechanisms to iteratively generate high-quality rewrites.   \textsc{LLM-R\textsuperscript{2}}~\cite{li2024llm} enhances a rule-based query rewriting system by using LLMs to suggest rewrite rules guided by contrastive demonstrations.
Despite variations in the quality and quantity of their rules,  rule-based techniques are fundamentally incapable of
optimizing new query patterns that have not been seen before. Though synthesis-based approaches~\cite{dong2023slabcity} are not constrained by rules,  they struggle with complex queries, such as those in the TPC-DS benchmark, which require navigating an enormously large search space. To address these limitations, more recent work has explored rewriting beyond rule-based boundaries by leveraging LLMs more directly. \revone{LITHE~\cite{llm-rw2} prompts an LLM with one selected natural language rule chosen from six handcrafted rewrite rules, each accompanied by an example, to guide the rewrite for a given query. Because this rule set is fixed, LITHE's knowledge cannot grow and evolve with more database feedback over time. QUITE~\cite{song2025quite} models the query rewrite process as a Markov Decision Process and at each timestep it chooses a refinement action (e.g., join reordering, predicate pushdown) by matching the query against a curated knowledge base built from database documentation and community sources. This design constrains the search to known actions and documented heuristics. In contrast, \tool is not restricted to a fixed action library: it can propose novel rewrites and accumulate generalizable rewrite knowledge (NLR2s) over time, including patterns that are not present in existing literature.}

\head{LLMs in databases} Beyond query rewriting, LLMs support NL2SQL with strong Spider results~\cite{li2023can};  schema matching by inferring column correspondences~\cite{parciak2407schema}; data augmentation with SQLsynth~\cite{tian2025text}; \revthree{equivalence checking via LLM\mbox{-}SQL\mbox{-}Solver with pragmatic “relaxed” equivalence~\cite{zhao2023llm};} tuning/diagnostics via DB\mbox{-}BERT and D\mbox{-}Bot~\cite{trummer2022db,zhou2024d}; and broader DBMS integration for optimization and indexing in DB\mbox{-}GPT~\cite{zhou2023dbgpt}. Collectively, these works position LLMs as versatile assistants across the database stack.
\vspace{-1mm}

\section{Conclusion}
\label{sec:conclusion}

This paper presents \tool, a holistic system that leverages LLMs for query rewriting beyond rules. We introduce  Natural Language Rewrite Rules (NLR2s) for transferring knowledge from rewriting one query to another, and a novel counterexample-guided technique that iteratively corrects the syntactic and semantic errors in the rewrites. Our empirical analysis demonstrates that \tool optimize a wider range of complex queries and achieve faster speedup than all baselines on the standard TPC-DS and JOB benchmarks and \revtwo{their SQLStorm-generated variants}.

% We evaluate \tool on the standard TPC-DS and JOB benchmarks and \revtwo{their SQLStorm-generated variants}. \tool consistently optimizes more queries at every speedup threshold than all baselines. \revtwo{At the $\geq$2x threshold on TPC-DS, \tool improves 25 queries—~1.35x more than LLM-driven baselines and ~2.6x more than LLM-enhanced rule-based baselines—and the gap widens further on TPC-DS (SQLStorm); on JOB and its SQLStorm variant, where queries are simpler, absolute gains are smaller but \tool still leads.

% \tool is now open-source but, recently, also implemented for commercial adoption by Keebo.~\footnote{\url{https://keebo.ai}}

% , significantly reducing the LLM costs and the manual effort required for verification. 
% \tool speeds up   22 out of 99 TPC queries (the most complex public benchmark) by more than 2x, which is  2.5x--3.2x higher coverage than state-of-the-art traditional query rewriting and 2.1x higher than the out-of-the-box LLM baseline.
% \input{sections/9_acknowledgement}

\clearpage

%\clearpage

\bibliographystyle{ACM-Reference-Format}
\bibliography{main.bib}

\clearpage

\begin{appendices}

\section{Representative rewrite patterns discovered by \tool}\label{app:pattern}

\begin{enumerate}[leftmargin=0.5cm]
\item  Pivot with conditional aggregation instead of UNION-then-self-join.
\item Replace multiple \texttt{EXISTS} conditions with a precomputed set via joins.
\item Rewrite \texttt{COUNT(*)>0} as \texttt{EXISTS} (aggregate $\rightarrow$ semi-join; enables short-circuiting).
\item Replace a bounded self-recursive CTE with a non-recursive base query \texttt{CROSS JOIN}ed to a small numbers table.
\item Lift a per-key summary (e.g., item-level totals) into one CTE and drive all branches from it.
\item Consolidate repetitive subqueries into a single aggregation subquery with conditional expressions.
\item Decorrelate \texttt{IN} by joining a \texttt{SELECT DISTINCT} key,... mapping table instead of re-checking the subquery. 
\end{enumerate}

\subsection{Pattern 1}

\head{Pivot with conditional aggregation instead of UNION-then-self-join}

\begin{listing}[h]
  \tiny
  \centering
  % BEFORE
  \begin{minipage}[t]{0.49\linewidth}
    \begin{tcolorbox}[colback=white, colframe=black, boxsep=2mm, left=4pt, right=4pt, top=0.5mm, bottom=0.5mm]
\begin{minted}[
  frame=none,
  framesep=1.5mm,
  numbersep=2mm,
  baselinestretch=1.0,
  fontsize=\tiny,
  style=bw,
  breaklines,
  breakanywhere
]{sql}
-- BEFORE: UNION of slices + multi-self-join
WITH slices AS (
  SELECT
    cust_id,
    yr,
    'S' AS ch,
    MAX(val) AS total
  FROM store
  WHERE yr IN (1999, 2000)
  GROUP BY
    cust_id, yr

  UNION ALL

  SELECT
    cust_id,
    yr,
    'W' AS ch,
    MAX(val) AS total
  FROM web
  WHERE yr IN (1999, 2000)
  GROUP BY
    cust_id, yr
)
SELECT
  s2.cust_id
FROM
  slices s1,
  slices s2,
  slices w1,
  slices w2
WHERE
  s1.ch = 'S' AND s1.yr = 1999 AND
  s2.ch = 'S' AND s2.yr = 2000 AND
  w1.ch = 'W' AND w1.yr = 1999 AND
  w2.ch = 'W' AND w2.yr = 2000 AND
  s1.total > 0 AND
  w1.total > 0 AND
  (w2.total / w1.total) >
  (s2.total / s1.total)
LIMIT 100;
\end{minted}
    \end{tcolorbox}
    \vspace{-1mm}
    \caption*{(a) Original}
  \end{minipage}
  \hfill
  % AFTER
  \begin{minipage}[t]{0.49\linewidth}
    \begin{tcolorbox}[colback=white, colframe=black, boxsep=2mm, left=4pt, right=4pt, top=0.5mm, bottom=0.5mm]
\begin{minted}[
  frame=none,
  framesep=1.5mm,
  numbersep=2mm,
  baselinestretch=1.0,
  fontsize=\tiny,
  style=bw,
  breaklines,
  breakanywhere
]{sql}
-- AFTER: conditional-agg pivot + one join
WITH s AS (
  SELECT
    cust_id,
    MAX(CASE WHEN yr = 1999
      THEN val END) AS s_1999,
    MAX(CASE WHEN yr = 2000
      THEN val END) AS s_2000
  FROM store
  WHERE yr IN (1999, 2000)
  GROUP BY cust_id
),
w AS (
  SELECT
    cust_id,
    MAX(CASE WHEN yr = 1999
      THEN val END) AS w_1999,
    MAX(CASE WHEN yr = 2000
      THEN val END) AS w_2000
  FROM web
  WHERE yr IN (1999, 2000)
  GROUP BY cust_id
)
SELECT
  s.cust_id
FROM
  s JOIN w USING (cust_id)
WHERE
  s.s_1999 > 0 AND
  w.w_1999 > 0 AND
  (w.w_2000 / w.w_1999) >
  (s.s_2000 / s.s_1999)
LIMIT 100;
\end{minted}
    \end{tcolorbox}
    \vspace{-1mm}
    \caption*{(b) Rewritten}
  \end{minipage}
  \vspace{-1mm}
  \captionsetup{font=small,name=Listing}
  \caption{Replace UNION\,+\,multi-self-join over per-slice aggregates with per-channel conditional aggregates and a single join.}
  \inv
  \label{listing:pivot-union-selfjoin}
\end{listing}

Replace a UNION of per-slice (year, channel) aggregates plus downstream self-joins with per-channel conditional aggregates (pivoted columns) computed once, then compare the resulting measures with a single join.

\head{Idea-level matches}
\begin{enumerate}
    \item Super-operators collapse multiple filtered aggregates and their UNION ALL into a single fused computation to avoid extra stages/shuffles—conceptually the same ``multi-branch aggregation fusion'' you apply per channel. \cite{leeka2019incorporating}
\item Athena fusion removes UNION ALL by merging multiple branches over the same subquery so the expensive block is computed once, then fanned out (they tag branches instead of pivoting to columns). \cite{bruno2022computation}
\end{enumerate}

\head{Why no exact match}
Prior work shows explicit UNION ALL elimination via fusion (and super-operator stage collapse), but they do not present the specific transformation that pivots the per-slice aggregates into columns (e.g., s\_1999, s\_2000, w\_1999, w\_2000) to avoid the downstream multi-self-join and express the cross-slice ratio predicate directly.

\subsection{Pattern 2}

\head{Replace multiple \texttt{EXISTS} conditions with a precomputed set via joins}

\begin{listing}[h]
  \tiny
  \centering
  % BEFORE
  \begin{minipage}[t]{0.49\linewidth}
    \begin{tcolorbox}[colback=white, colframe=black, boxsep=2mm, left=4pt, right=4pt, top=0.5mm, bottom=0.5mm]
\begin{minted}[
  frame=none,
  framesep=1.5mm,
  numbersep=2mm,
  baselinestretch=1.0,
  fontsize=\tiny,
  style=bw,
  breaklines,
  breakanywhere
]{sql}
-- BEFORE: multi-EXISTS channel gating
SELECT
  cd.cd_gender,
  ca.ca_state,
  COUNT(*) AS cnt
FROM customer       c
JOIN customer_demographics cd
  ON cd.cd_demo_sk = c.c_current_cdemo_sk
JOIN customer_address ca
  ON ca.ca_address_sk = c.c_current_addr_sk
WHERE EXISTS (
  SELECT 1
  FROM store_sales s
  JOIN date_dim d
    ON s.ss_sold_date_sk = d.d_date_sk
  WHERE s.ss_customer_sk = c.c_customer_sk
    AND d.d_year = 2001
)
AND (
  EXISTS (
    SELECT 1
    FROM web_sales w
    JOIN date_dim d
      ON w.ws_sold_date_sk = d.d_date_sk
    WHERE w.ws_bill_customer_sk = c.c_customer_sk
      AND d.d_year = 2001
  )
  OR EXISTS (
    SELECT 1
    FROM catalog_sales cs
    JOIN date_dim d
      ON cs.cs_sold_date_sk = d.d_date_sk
    WHERE cs.cs_ship_customer_sk = c.c_customer_sk
      AND d.d_year = 2001
  )
)
GROUP BY cd.cd_gender, ca.ca_state
LIMIT 100;
\end{minted}
    \end{tcolorbox}
    \vspace{-1mm}
    \caption*{(a) Original}
  \end{minipage}
  \hfill
  % AFTER
  \begin{minipage}[t]{0.49\linewidth}
    \begin{tcolorbox}[colback=white, colframe=black, boxsep=2mm, left=4pt, right=4pt, top=0.5mm, bottom=0.5mm]
\begin{minted}[
  frame=none,
  framesep=1.5mm,
  numbersep=2mm,
  baselinestretch=1.0,
  fontsize=\tiny,
  style=bw,
  breaklines,
  breakanywhere
]{sql}
-- AFTER: one eligibility set + one join
WITH s AS (
  SELECT ss_customer_sk AS cust_sk
  FROM store_sales s
  JOIN date_dim d
    ON s.ss_sold_date_sk = d.d_date_sk
  WHERE d.d_year = 2001
  GROUP BY ss_customer_sk
),
w AS (
  SELECT ws_bill_customer_sk AS cust_sk
  FROM web_sales w
  JOIN date_dim d
    ON w.ws_sold_date_sk = d.d_date_sk
  WHERE d.d_year = 2001
  GROUP BY ws_bill_customer_sk
),
c AS (
  SELECT cs_ship_customer_sk AS cust_sk
  FROM catalog_sales cs
  JOIN date_dim d
    ON cs.cs_sold_date_sk = d.d_date_sk
  WHERE d.d_year = 2001
  GROUP BY cs_ship_customer_sk
),
qc AS (  -- qualified customers: Store AND (Web OR Catalog)
  SELECT s.cust_sk
  FROM s
  JOIN (
    SELECT cust_sk FROM w
    UNION
    SELECT cust_sk FROM c
  ) u
    ON s.cust_sk = u.cust_sk
)
SELECT
  cd.cd_gender,
  ca.ca_state,
  COUNT(*) AS cnt
FROM customer c
JOIN qc
  ON c.c_customer_sk = qc.cust_sk
JOIN customer_demographics cd
  ON cd.cd_demo_sk = c.c_current_cdemo_sk
JOIN customer_address ca
  ON ca.ca_address_sk = c.c_current_addr_sk
GROUP BY cd.cd_gender, ca.ca_state
LIMIT 100;
\end{minted}
    \end{tcolorbox}
    \vspace{-1mm}
    \caption*{(b) Rewritten}
  \end{minipage}
  \vspace{-1mm}
  \captionsetup{font=small,name=Listing}
  \caption{Replace multiple \texttt{EXISTS} predicates with a single
  precomputed per-customer eligibility set (via UNION/INTERSECT) and
  join to that set once.}
  \inv
  \label{listing:exists-to-eligibility-set}
\end{listing}

Replace multiple \texttt{EXISTS} predicates with a single precomputed eligibility set and join to that set once.

\head{Idea-level matches}
\begin{enumerate}
  \item Calcite decorrelates a single \texttt{EXISTS} into a semi-join,
  but does not factor multiple \texttt{EXISTS} into one shared set.
  \item GSQF/Athena fuse repeated subplans to avoid redundant scans;
  here we express boolean gating as UNION/INTERSECT over keys to enable
  one reusable join.
\end{enumerate}

\head{Why no exact match}
Prior work shows semi-join decorrelation (single \texttt{EXISTS}) and
fusion of identical or same-source subplans, but not the specific
boolean-to-set factoring that consolidates several correlated
\texttt{EXISTS} across different fact tables into one eligibility CTE
and a single join.

\subsection{Pattern 3}

\head{Rewrite \texttt{COUNT(*)>0} as \texttt{EXISTS} (aggregate $\rightarrow$ semi-join; enables short-circuiting)}

\begin{listing}[h]
  \tiny
  \centering
  % BEFORE
  \begin{minipage}[t]{0.49\linewidth}
    \begin{tcolorbox}[colback=white, colframe=black, boxsep=2mm,
      left=4pt, right=4pt, top=0.5mm, bottom=0.5mm]
\begin{minted}[
  frame=none,
  framesep=1.5mm,
  numbersep=2mm,
  baselinestretch=1.0,
  fontsize=\tiny,
  style=bw,
  breaklines,
  breakanywhere
]{sql}
-- BEFORE: scalar COUNT(*) > 0 in a
-- correlated subquery
SELECT DISTINCT p.name
FROM product p
WHERE p.mfg_id BETWEEN 700 AND 740
  AND (
    SELECT COUNT(*)
    FROM product q
    WHERE q.mfg = p.mfg
      AND (
        (q.cat = 'Women' AND
         q.color IN ('forest','lime'))
        OR
        (q.cat = 'Men'   AND
         q.color IN ('sky','powder'))
      )
  ) > 0
ORDER BY p.name
LIMIT 100;
\end{minted}
    \end{tcolorbox}
    \vspace{-1mm}
    \caption*{(a) Original}
  \end{minipage}
  \hfill
  % AFTER
  \begin{minipage}[t]{0.49\linewidth}
    \begin{tcolorbox}[colback=white, colframe=black, boxsep=2mm,
      left=4pt, right=4pt, top=0.5mm, bottom=0.5mm]
\begin{minted}[
  frame=none,
  framesep=1.5mm,
  numbersep=2mm,
  baselinestretch=1.0,
  fontsize=\tiny,
  style=bw,
  breaklines,
  breakanywhere
]{sql}
-- AFTER: EXISTS (semi-join form),
-- no aggregate in the subquery
SELECT DISTINCT p.name
FROM product p
WHERE p.mfg_id BETWEEN 700 AND 740
  AND EXISTS (
    SELECT 1
    FROM product q
    WHERE q.mfg = p.mfg
      AND (
        (q.cat = 'Women' AND
         q.color IN ('forest','lime'))
        OR
        (q.cat = 'Men'   AND
         q.color IN ('sky','powder'))
      )
  )
ORDER BY p.name
LIMIT 100;
\end{minted}
    \end{tcolorbox}
    \vspace{-1mm}
    \caption*{(b) Rewritten}
  \end{minipage}
  \vspace{-1mm}
  \captionsetup{font=small,name=Listing}
  \caption{Replace a correlated scalar subquery
  with \texttt{COUNT(*)>0} by an \texttt{EXISTS}
  predicate (semi-join form).}
  \label{listing:count-exists}
\end{listing}

% One-line rule summary
Rewrite correlated \texttt{COUNT(*)>0} subqueries as \texttt{EXISTS}, eliminating the aggregate and enabling a semi-join evaluation.

\head{Idea-level matches}
\begin{enumerate}[leftmargin=0.5cm,itemsep=1pt,topsep=2pt]
  \item Calcite-style subquery removal/decorrelation (convert subqueries to joins/correlates), plus boolean simplification/constant folding.
\end{enumerate}

\head{Why no exact match}
Calcite does not expose a dedicated “\texttt{COUNT(*)>0} $\rightarrow$ \texttt{EXISTS}” rule; its subquery-removal and decorrelation pipeline does not yield this specific rewrite pattern.

\begin{comment}
    
\end{comment}

\newpage

\subsection{Pattern 4}

\head{Replace a bounded self-recursive CTE with a non-recursive base query \texttt{CROSS JOIN}ed to a small numbers table}

\begin{listing}[h]
  \tiny
  \centering
  % BEFORE
  \begin{minipage}[t]{0.49\linewidth}
    \begin{tcolorbox}[colback=white, colframe=black, boxsep=2mm,
      left=4pt, right=4pt, top=0.5mm, bottom=0.5mm]
\begin{minted}[
  frame=none,
  framesep=1.5mm,
  numbersep=2mm,
  baselinestretch=1.0,
  fontsize=\tiny,
  style=bw,
  breaklines,
  breakanywhere
]{sql}
-- BEFORE: bounded recursion (rn < 5)
WITH RECURSIVE r AS (
  SELECT
    id,
    city,
    ROW_NUMBER() OVER (
      PARTITION BY city
      ORDER BY id
    ) AS rn
  FROM addr
  WHERE state IN ('CA','NY')

  UNION ALL

  SELECT
    id,
    city,
    rn + 1
  FROM r
  WHERE rn < 5
)
SELECT
  id, city, rn
FROM r
LIMIT 100;
\end{minted}
    \end{tcolorbox}
    \vspace{-1mm}
    \caption*{(a) Original}
  \end{minipage}
  \hfill
  % AFTER
  \begin{minipage}[t]{0.49\linewidth}
    \begin{tcolorbox}[colback=white, colframe=black, boxsep=2mm,
      left=4pt, right=4pt, top=0.5mm, bottom=0.5mm]
\begin{minted}[
  frame=none,
  framesep=1.5mm,
  numbersep=2mm,
  baselinestretch=1.0,
  fontsize=\tiny,
  style=bw,
  breaklines,
  breakanywhere
]{sql}
-- AFTER: numbers table * base (no recursion)
WITH base AS (
  SELECT
    id,
    city,
    ROW_NUMBER() OVER (
      PARTITION BY city
      ORDER BY id
    ) AS rn
  FROM addr
  WHERE state IN ('CA','NY')
),
nums(n) AS (
  VALUES (1),(2),(3),(4),(5)
)
SELECT
  b.id,
  b.city,
  (b.rn + n - 1) AS rn
FROM base AS b
CROSS JOIN nums
WHERE (b.rn + n - 1) <= 5
LIMIT 100;
\end{minted}
    \end{tcolorbox}
    \vspace{-1mm}
    \caption*{(b) Rewritten}
  \end{minipage}
  \vspace{-1mm}
  \captionsetup{font=small,name=Listing}
  \caption{Eliminate bounded recursion by CROSS JOINing the base
  with a small numbers table to generate the expanded rows.}
  \inv
  \label{listing:bounded-recursion-numbers}
\end{listing}

\head{Idea-level matches}
\begin{enumerate}[leftmargin=0.55cm,itemsep=1pt,parsep=0pt,topsep=2pt]
  \item None among \cite{sarthi2020generalized,leeka2019incorporating,
  modi2021new,bruno2022computation}: these target fusion/reuse or
  shuffle reduction, not recursion elimination via enumerators.
\end{enumerate}

\head{Why no exact match}
Prior work does not present the specific transformation of a bounded
self-recursive CTE into a non-recursive base query CROSS JOINed with
a small numbers table.

\newpage

\subsection{Pattern 5}

\head{Lift a per-key summary (e.g., item-level totals) into one CTE and drive all branches from it}

\begin{listing}[h]
  \tiny
  \centering
  % BEFORE
  \begin{minipage}[t]{0.49\linewidth}
    \begin{tcolorbox}[colback=white, colframe=black, boxsep=2mm, left=4pt, right=4pt, top=0.5mm, bottom=0.5mm]
\begin{minted}[
  frame=none,
  framesep=1.5mm,
  numbersep=2mm,
  baselinestretch=1.0,
  fontsize=\tiny,
  style=bw,
  breaklines,
  breakanywhere
]{sql}
-- BEFORE: repeated scans of fact table
WITH rs AS (
  SELECT
    item_id,
    SUM(qty)  AS tot_qty,
    SUM(price * qty) AS tot_sales
  FROM sales
  GROUP BY item_id
),
top AS (
  SELECT
    r.item_id,
    r.tot_qty,
    r.tot_sales
  FROM rs r
  WHERE
    RANK() OVER (
      PARTITION BY r.item_id
      ORDER BY r.tot_sales DESC
    ) <= 10
)
SELECT
  t.item_id,
  t.tot_qty,
  t.tot_sales
FROM top t
WHERE t.tot_sales IS NOT NULL

UNION ALL

SELECT
  t.item_id,
  NULL AS tot_qty,
  NULL AS tot_sales
FROM top t
WHERE NOT EXISTS (
  SELECT 1
  FROM sales s
  WHERE s.item_id = t.item_id
);
\end{minted}
    \end{tcolorbox}
    \vspace{-1mm}
    \caption*{(a) Original}
  \end{minipage}
  \hfill
  % AFTER
  \begin{minipage}[t]{0.49\linewidth}
    \begin{tcolorbox}[colback=white, colframe=black, boxsep=2mm, left=4pt, right=4pt, top=0.5mm, bottom=0.5mm]
\begin{minted}[
  frame=none,
  framesep=1.5mm,
  numbersep=2mm,
  baselinestretch=1.0,
  fontsize=\tiny,
  style=bw,
  breaklines,
  breakanywhere
]{sql}
-- AFTER: one per-key summary reused everywhere
WITH s AS (
  SELECT
    item_id,
    SUM(qty)  AS tot_qty,
    SUM(price * qty) AS tot_sales
  FROM sales
  GROUP BY item_id
),
top AS (
  SELECT
    s.item_id,
    s.tot_qty,
    s.tot_sales
  FROM s
  WHERE
    RANK() OVER (
      PARTITION BY s.item_id
      ORDER BY s.tot_sales DESC
    ) <= 10
),
pref AS (
  SELECT
    t.item_id,
    t.tot_qty,
    t.tot_sales
  FROM top t
  WHERE t.tot_sales IS NOT NULL
)
SELECT
  p.item_id,
  p.tot_qty,
  p.tot_sales
FROM pref p

UNION ALL

SELECT
  t.item_id,
  NULL AS tot_qty,
  NULL AS tot_sales
FROM top t
WHERE NOT EXISTS (
  SELECT 1
  FROM s
  WHERE s.item_id = t.item_id
);
\end{minted}
    \end{tcolorbox}
    \vspace{-1mm}
    \caption*{(b) Rewritten}
  \end{minipage}
  \vspace{-1mm}
  \captionsetup{font=small,name=Listing}
  \caption{Lift a per-key summary into one CTE and drive all branches from it, eliminating rescans of the fact table.}
  \inv
  \label{listing:perkey-summary-reuse}
\end{listing}

\head{Idea-level matches}
\begin{enumerate}
  \item Computation reuse: compute a shared sub-block once and fan it out to consumers (\emph{Athena} fusion). 
  \item Classic algebraic rewrites: common subexpression sharing and aggregate-before-join (Calcite rule family).
\end{enumerate}

\head{Why no exact match}
Prior work endorses sharing/fusion in spirit, but does not present a rule that \emph{enforces} a single per-key summary as the canonical input for \emph{all} branches without additional tag-and-merge machinery.

\clearpage

\section{NLR2 $\to$ guarded executable rules}\label{app:extract_rule}

In the context of recurring query execution, \tool can serve as an automated system to propose query pairs that uncover valuable pattern-matching rewrite rules. \tool is not designed to replace rule-based methods; rather, it leverages the capabilities of LLMs to identify promising rules that can be integrated into existing database. \tool returns verified  query pairs that deliver measured speedups together with the associated NLR2. A rule distilled from only one such pair is brittle and often fails to generalize—it overfits incidental syntax choices and misses necessary preconditions. We can aggregate multiple verified pairs tied to the same NLR2, and extract the common AST/plan transformation plus explicit guards/preconditions. As more pairs accumulate, the resulting pattern-matching rule becomes more general  while remaining safe.

% ---------- Pivot-fusion rewrite rule (copy-paste ready) ----------
% \subsubsection{Pivot-fusion rule (UNION + multi-self-join $\rightarrow$ conditional aggregates + single join)}
\begin{tcolorbox}[colback=white,colframe=black,boxsep=1mm,left=2pt,right=2pt,top=1mm,bottom=1mm]
\footnotesize
\textbf{Intent.} Replace a \emph{UNION} of per-slice aggregates plus downstream self-joins with per-channel
\emph{conditional aggregates} (pivoted columns), then use a single join across channels to express cross-slice predicates.

\medskip
\textbf{Preconditions.}
\begin{enumerate}\itemsep2pt
  \item Each UNION arm groups by the same keys $G$ (e.g., \texttt{cust\_id}) and produces one measure per $(G,\text{slice})$.
  \item The slice domain is finite (e.g., \texttt{year IN (1999,2000)}).
  \item Downstream self-joins of the UNION only select particular $(\text{channel},\text{slice})$ rows and compare functions of their measures.
  \item Aggregator is NULL-ignoring (e.g., \texttt{SUM}, \texttt{MAX}, \texttt{MIN}, \texttt{COUNT}), so \texttt{AGG(CASE WHEN slice=$\sigma$ THEN m END)} equals the per-slice aggregate.
\end{enumerate}

\textbf{Rewrite.} For each channel $c$:
\[
\begin{aligned}
\texttt{SELECT } & G, \\
& \big\{\,\texttt{AGG(CASE WHEN slice = } \sigma_j
   \texttt{ THEN m END) AS } v_{c,\sigma_j}\,\big\}_{j} \\
\texttt{FROM } & \texttt{... GROUP BY } G
\end{aligned}
\]

Then join these per-channel results on $G$ and rewrite predicates using the pivoted columns $v_{c,\sigma_j}$.

\textbf{Correctness (sketch).} The UNION has exactly one row per $(G,c,\sigma)$; the rewrite creates one column per $(c,\sigma)$.
NULL-ignoring aggregation preserves per-slice values; self-join filters become column selections; equality joins on $G$ remain.
\end{tcolorbox}

\FloatBarrier % optional: keep prior floats from jumping over
\begin{table}[H]
\footnotesize
\centering
\resizebox{\columnwidth}{!}{%
\begin{tabular}{l l l l}
\toprule
 & TPC-DS Q74 Pair & TPC-DS Q4 Pair & Note \\
\midrule
\# channels & 2 (S, W) & 3 (S, C, W) & Scales to $k\ge2$ \\
Aggregator & MAX & SUM & Works for NULL-ignoring AGGs \\
Measure expr & net\_paid & composite & Syntax-agnostic \\
Slice set & \{1999,2000\} & \{1999,2000\} & Any finite set \\
Downstream logic & 1 ratio & 2 ratios & Multiple predicates OK \\
Passthrough attrs & names & names+country & Preserved \\
\bottomrule
\end{tabular}}
\captionsetup{font=small,name=Table}
\caption{Evidence the rule generalizes across channels, aggregators, and measures.}
\label{tab:pivot-fusion-evidence}
\end{table}

% \begin{table*}[tbp]
% \footnotesize
% \centering
% \setlength{\tabcolsep}{4pt}
% \def\arraystretch{1.05}
% \begin{tabular}{l l l l}
% \toprule
%  & TPC-DS Q74 Pair & TPC-DS Q4 Pair & Note \\
% \midrule
% \# channels & 2 (S, W) & 3 (S, C, W) & Scales to $k\ge2$ \\
% Aggregator & MAX & SUM & Works for NULL-ignoring AGGs \\
% Measure expr & Simple (\texttt{net\_paid}) & Composite arithmetic & Syntax-agnostic \\
% Slice set & \{1999, 2000\} & \{1999, 2000\} & Any finite set \\
% Downstream logic & 1 ratio & 2 ratios (C vs S, C vs W) & Multiple predicates OK \\
% Passthrough attrs & Names & Names + country & Preserved via grouping \\
% \bottomrule
% \end{tabular}
% \captionsetup{font=small,name=Table}
% \caption{Evidence that the rule generalizes beyond a single pair: different channels, aggregators, and measures.}
% \label{tab:pivot-fusion-evidence}
% \end{table*}

% ---------- End pivot-fusion snippet ----------

Pattern~1 presented in Appendix~\ref{app:pattern} applies to three standard TPC-DS queries: Q4, Q11, Q74.
Across all three, \tool can successfully rewrite them with the guidance of the same NLR2. Considering these pairs jointly lets us
(1) factor out query-specific quirks, (2) state explicit preconditions (shared grouping
keys, finite slice domain, NULL-ignoring aggregates), and (3) codify a single guarded,
executable rule that is both more general and safe to apply.

 \end{appendices}

\end{document}